\documentclass[12pt,a4paper]{article}

\usepackage{amssymb,amsmath,eucal}
\usepackage[dvips]{lscape,graphicx}
\usepackage{cite}

\newcommand{\bc}{\begin{center}}
\newcommand{\ec}{\end{center}}
\newcommand{\bd}{\begin{displaymath}}
\newcommand{\ed}{\end{displaymath}}
\newcommand{\be}{\begin{equation}}
\newcommand{\ee}{\end{equation}}
\newcommand{\ba}{\begin{array}}
\newcommand{\ea}{\end{array}}
\newcommand{\bt}{\begin{tabular}}
\newcommand{\et}{\end{tabular}}
\newcommand{\un}{\underline}
\newcommand{\ov}{\overline}

\newcommand{\ct}{\cite}

\newcommand{\lb}{\label}
\newcommand{\bp}{\begin{picture}}
\newcommand{\ep}{\end{picture}}
\newcommand{\bfi}{\begin{figure}}
\newcommand{\plaqr}{$\biggl(\quad\raisebox{-2pt}{\mbox{\framebox(0,12){\phantom{a}}\hspace{-5.6mm}
\dashbox{3}(12,12)[b]{\phantom{a}}}}\biggr)$}
\newcommand{\plaql}{$\biggl(\raisebox{-2pt}{\mbox{\framebox(0,12){\phantom{a}}\hspace{-1.3mm}
\dashbox{3}(12,12)[b]{\phantom{a}}}}\biggr)$}
\newcommand{\plaqt}{$\biggl(\raisebox{-2pt}{\mbox{\raisebox{12pt}{\framebox(11,0){\phantom{a}}}\hspace{-5.5mm}
\dashbox{3}(12,12)[b]{\phantom{a}}}}\biggr)$}
\newcommand{\plaqb}{$\biggl(\raisebox{-2pt}{\mbox{\framebox(11,0)
{\phantom{a}}\hspace{-5.5mm}
\dashbox{2}(12,12)[b]{\phantom{a}}}}\biggr)$}

\newcommand{\link}{\begin{array}{l}\begin{picture}(22,4)
    \put(0,2.5){\circle*{4}}
    \put(20,2.5){\circle*{4}}
    \put(0,2.5){\line(1,0){20}}
\end{picture}\end{array}}

\begin{document}

\vspace{1cm}

\title{\huge {\bf{ Monopoles near the Planck Scale
and Unification }}}
\author{{\Large L.V.Laperashvili\thanks{laper@heron.itep.ru}, D.A.Ryzhikh\thanks{ryzhikh@heron.itep.ru}}\\
\it Institute of Theoretical and Experimental Physics,\\
\it B. Cheremushkinskaya 25, 117218 Moscow, Russia\\
and\\
{\Large H.B.Nielsen\thanks{hbech@mail.desy.de, hbech@nbi.dk}}\\
\it DESY, Notkestrasse 85, 22603 Hamburg,\\
\it Niels Bohr Institute, DK-2100, Copenhagen, Denmark}

\date{}

\maketitle
\thispagestyle{empty}
\vspace{1cm}
\newpage

\begin{abstract}

Considering our (3 + 1)-dimensional space-time as, in some way, discrete or
lattice with a parameter $a=\lambda_P$, where $\lambda_P$ is the Planck
length, we have investigated the additional contributions of lattice
artifact monopoles to beta-functions of the renormalisation group equations
for the running fine structure constants $\alpha_i(\mu)$ (i=1,2,3
correspond to the U(1), SU(2) and SU(3) gauge groups of the Standard Model)
in the Family Replicated Gauge Group Model (FRGGM) which is an extension of
the Standard Model at high energies. It was shown that monopoles have
$N_{fam}$ times smaller magnetic charge in FRGGM than in SM
($N_{fam}$ is the number of families in FRGGM).
We have estimated also the enlargement of a number of
fermions in FRGGM leading to the suppression of the asymptotic freedom in
the non--Abelian theory. We have shown that, in contrast to the case
of AntiGUT when the FRGGM undergoes the breakdown at $\mu=\mu_G\sim
10^{18}\,$ GeV, we have the possibility of unification if the
FRGGM--breakdown occurs at $\mu_G\sim 10^{14}\,$ GeV.
By numerical calculations we obtained an example of the unification
of all gauge interactions (including gravity) at the scale
$\mu_{GUT}\approx 10^{18.4}$ GeV. We discussed the possibility of
$[SU(5)]^3$ or $[SO(10)]^3$ (SUSY or not SUSY) unifications.

\end{abstract}

\pagenumbering{arabic}

\newpage

\section{Introduction}

Trying to look insight the Nature and considering the physical processes
at very small distances, physicists have made attempts to explain the
well--known laws of low--energy physics as a consequence of the more
fundamental laws of Nature. The contemporary physics of the electroweak
and strong interactions is described by the Standard Model (SM) which
combines the Glashow--Salam--Weinberg electroweak theory with QCD ---
theory of strong interactions.

The gauge group of symmetry in SM is:.
\begin{equation}
   SMG = SU(3)_c\times SU(2)_L\times U(1)_Y,            \lb{1}
\end{equation}
which describes the present elementary particle physics up to the scale
$\approx 100$ GeV.

Having an interest in the fundamental laws of physics, we can consider
the two possibilities:

1. At the very small (Planck length) distances {\un{our space--time is
continuous}} and there exists the fundamental theory (maybe with a
very high symmetry) which we do not know at present time.

2. At the very small distances {\un{our space--time is
discrete}}, and this discreteness influences on the Planck scale physics.

The item 2 is a base of Random Dynamics (RD)
which was suggested and developed in Refs.\ct{2a}-\ct{2m} as a
theory of physical processes proceeding at small distances of order of the
Planck scale $\lambda_P=M_{Pl}^{-1}$:
\begin{equation}
                M_{Pl}=1.22\cdot 10^{19}\,{\mbox{GeV}}.     \lb{2}
\end{equation}
The theory of Scale Relativity (SR) \ct{10na}-\ct{10nc} is also related with
the item 2 and has a lot in common with RD.
SR assumes that the resolution of experimental measurements plays in quantum
physics a completely new role with respect of the classical theory, and
the space--time is described by a metric based on generalized, explicitly
scale--dependent, metric potentials:
\begin{equation}
g_{\mu\nu} = g_{\mu\nu}(t,x,y,z;\,\Delta t,\,\Delta x,\,\Delta y,\,\Delta z).
                                          \lb{3}
\end{equation}
SR theory predicts that there exists a minimal scale of the space--time
resolution equal to the Planck length $\lambda_P$, which can be
considered as a fundamental scale of our Nature. This gives us a
reason to make an assumption that {\it our (3+1)--dimensional space is
discrete on the fundamental level.}

This is an initial (basic) point of view of the present theory,
i.e. we take a discreteness as existing, not as the lattice
computation trick in QCD, say.
In the simplest case we can imagine our (3+1) space--time as a
regular hypercubic lattice with a parameter $a=\lambda_P$. Then
the lattice artifact monopoles can play an essential role near the
Planck scale. But, of course, it is necessary to comment that we
do not know (at least, on the level of our today knowledge), what
lattice--like structure (random lattice, or foam, or string
lattice, etc.) is realized in the description of physical processes
at very small distances \ct{11}, even if there should be a lattice.

The main aim of the present paper is to show that monopoles cannot be seen
in the usual SM up to the Planck scale, because they have a huge magnetic
charge and are completely confined. Supersymmetric extensions of SM,
known in literature, does not help to see monopoles. We suggest to consider
a possibility of the existence of monopoles in our World
by extending the Standard Model Group (SMG) to the
Family Replicated Gauge Group ${(SMG)}^{N_{fam}}$, where $N_{fam}$ is the
number of families of quarks and leptons. This Family Replicated Gauge
Group (FRGG) was very successful in describing the SM parameters: the
authors of Refs.\ct{35}--\ct{38}, using the breakdown of FRGG to SM, have
found a very good fit to conventional experimental data for all fermion
masses and mixing angles in SM, also explained the modern experiments for
neutrino oscillations.

In the present paper we have written the renormalisation group equations
(RGEs) for electric and magnetic fine structure constants (Section 2)
and investigated the evolution of gauge and gravitational fine
structure constants in Section 3.

The dropping (diminishing) of the monopole charges in FRGG
was considered in Section 4. It was shown that if one
wants to see monopoles in Nature, it is necessary to drive in the direction of
FRGGs.

Multiple Point Principle (MPP) --- the existence of the Multiple Critical
Point (MCP) at the Planck scale --- is reviewed in Section 5.

Section 6 is devoted to the lattice results for the phase transition coupling
constants.

We have described a specific evolution of the U(1) fine structure constant in
SM and FRGGM in the presence of the lattice artifact monopoles, or
fundamental Higgs magnetic fields (Section 7), confirming the idea of MPP.

In Section 8 we have presented the predictions of FRGGM for the Planck scale
values of gauge fine structure constants. We show that AntiGUT exists in the
case of the breakdown of FRGG at $\mu_G \sim 10^{18}$ GeV (where $\mu $ is the
energy scale).

In Section 9 we have considered a new possibility of unification of all
interactions, including gravity, at the point $\mu_{GUT}=10^{18.4}$ GeV and
$\alpha_{GUT}^{-1}=27$, if the breakdown of FRGG occurs at
$\mu_G \sim 10^{14}$ GeV.

In Section 10 we have presented the discussion of two various scenarios
for the existence of Multiple Citical Point (MCP).

\section {Renormalisation Group Equations for Electric and Magnetic
Fine Structure Constants}

J.Schwinger was first \ct{41} who investigated the problem of renormalisation
of the magnetic charge in Quantum ElectroMagnetoDynamics (QEMD), i.e.
in the Abelian quantum field theory of electrically and magnetically
charged particles (with charges $e$ and $g$, respectively).

Considering the "bare" charges $e_0$ and $g_0$ and renormalised (effective)
charges $e$ and $g$, Schwinger (and later the authors of Refs.\ct{42} and
\ct{43}) obtained:
\begin{equation}
               e/g = e_0/g_0,              \lb{4a}
\end{equation}
what means the absence of the Dirac relation \ct{43a} for the renormalised
electric and magnetic charges.

But there exists another solution of this problem (see Refs.
\ct{44}-\ct{47} and review \ct{48}), which gives:
\begin{equation}
                       eg = e_0g_0 = 2\pi n,\quad n\in Z,
                                                        \lb{4b}
\end{equation}
i.e. the existence of the Dirac relation (charge quantization condition) for
both, bare and renormalised electric and magnetic charges.
Here we have $n=1$ for the minimal (elementary) charges.

These two cases lead to the two possibilities for the renormalisation group
equations (RGEs) describing the evolution of electric and magnetic fine
structure constants:
\begin{equation}
    \alpha = \frac{e^2}{4\pi}\quad{\mbox{and}}\quad
    \tilde \alpha = \frac{g^2}{4\pi},                   \lb{5}
\end{equation}
which obey the following RGEs containing the electric and magnetic
beta--functions:
\begin{equation}
\frac {d(\log \alpha(\mu))}{dt} = \pm \frac {d(\log \tilde \alpha(\mu))}{dt}
= \beta^{(e)}(\alpha) \pm \beta^{(m)}(\tilde \alpha).
                                                       \lb{6a}
\end{equation}
In Eq.(\ref{6a}) we have:
\begin{equation}
             t = \log(\frac{\mu^2}{\mu_R^2}),            \lb{6}
\end{equation}
where $\mu$ is the energy scale and $\mu_R$ is the renormalisation point.

The second possibility (with minuses) in Eq.(\ref{6a})
corresponds to the validity of the Dirac relation (\ref{4b}) for the
renormalised charges.  We believe only in this case considered by authors
in Ref.\ct{47} where we have used the Zwanziger formalism of QEMD
\ct{49},\ct{50}. In the present paper, excluding the Schwinger's
renormalisation condition (\ref{4a}), we assume only the Dirac relation
for running $\alpha$ and $\tilde \alpha$:
\be
      \alpha \tilde \alpha = \frac{1}{4}.        \lb{7}
\ee
It is necessary to comment that RGEs (\ref{6a}) are valid only for
$\mu > \mu_{threshold} = m_{mon}$, where $m_{mon}$ is the monopole mass.

In contrast to the method given in Ref.\ct{47}, there exists a simple way
to obtain Eq.(\ref{6a}) for single electric and magnetic charges of
the same type (scalar or fermionic).
The general expressions for RGEs are:
\begin{equation}
\frac {d(\log \alpha(\mu))}{dt} =
   \beta_1(\alpha) + \beta_2(\tilde \alpha) + C,
                                                           \lb{16A}
\end{equation}
\begin{equation}
\frac {d(\log \tilde \alpha(\mu))}{dt} =
\tilde \beta_1(\alpha) + \tilde \beta_2(\tilde \alpha) + \tilde C,
                                                           \lb{16B}
\end{equation}
The Dirac relation (\ref{7}) gives:
\begin{equation}
\frac {d(\log \alpha(\mu))}{dt} = -
\frac {d(\log \tilde \alpha(\mu))}{dt}.
                                                           \lb{16C}
\end{equation}
Using Eq. (\ref{16C}) and the duality symmetry of QEMD,
i.e. the symmetry under the interchange:
\begin{equation}
          \alpha \longleftrightarrow \tilde \alpha,
                                    \lb{16D}
\end{equation}
it is not difficult to obtain:
\begin{equation}
              C = \tilde C = 0,\quad
   \beta_1(\alpha) = -  \beta_2(\alpha) =
\tilde \beta_1(\alpha) = - \tilde \beta_2(\alpha) = \beta(\alpha),
                                                           \lb{16E}
\end{equation}
and we have the following RGE:
\begin{equation}
\frac {d(\log \alpha(\mu))}{dt} = - \frac {d(\log \tilde \alpha(\mu))}{dt} =
   \beta(\alpha) - \beta(\tilde \alpha).
                                                           \lb{16F}
\end{equation}
If monopole charges, together with electric ones, are sufficiently small,
then $\beta$--functions can be considered perturbatively:
\begin{equation}
   \beta(\alpha) = \beta_2 (\alpha /4\pi) + \beta_4 {(\alpha/4\pi)}^2 + ...
                                                         \lb{18x}
\end{equation}
and
\begin{equation}
   \beta(\tilde \alpha) = \beta_2 (\tilde \alpha /4\pi) +
         \beta_4 {(\tilde \alpha/4\pi)}^2 + ...
                                                        \lb{19x}
\end{equation}
with (see paper \ct{47} and references there):
\begin{equation}
     \beta_2 = \frac 13  \quad{\mbox{and}}\quad \beta_4 =1 \quad
-\quad {\mbox {for scalar particles}},
                           \lb{20x}
\end{equation}
and
\begin{equation}
     \beta_2 = \frac 43  \quad{\mbox{and}}\quad \beta_4 \approx 4
\quad-\quad {\mbox {for fermions}}.
                           \lb{21x}
\end{equation}
These cases were investigated in Ref.\ct{47}.
For scalar electric and magnetic charges we have \ct{47}:
\begin{equation}
   \frac {d(\log \alpha(\mu))}{dt} = - \frac {d(\log
   \tilde \alpha(\mu))}{dt} = \beta_2 \frac{\alpha - \tilde \alpha}{4\pi}(1 + 3
   \frac{\alpha + \tilde \alpha}{4\pi} + ...)
                                                      \lb{17x}
\end{equation}
with $\beta_2 = 1/3$, and approximately the same result is valid
for fermionic particles \ct{51} with $\beta_2 = 4/3$.
Eq.(\ref{17x}) shows that there exists a region when both fine structure
constants are perturbative. Approximately this region is given by the
following inequalities:
\begin{equation}
    0.2 \stackrel{<}{\sim }(\alpha, \tilde \alpha)
                    \stackrel{<}{\sim }1.
                                               \lb{22x}
\end{equation}
Using the Dirac relation (\ref{7}), we see from Eq.(\ref{17x})
that in the region (\ref{22x}) the two--loop contribution is not larger
than 30\% of the one--loop contribution, and the perturbation theory
can be realized in this case (see Refs.\ct{20p}-\ct{24p}).

It is necessary to comment that the region (\ref{22x}) almost coincides
with the region of phase transition couplings obtained in the lattice
U(1)--gauge theory (see Subsection 6.3).

\section{Evolution of Running Fine Structure Constants}

The usual definition of the SM coupling constants is given in
{\it the Modified minimal subtraction scheme}($\ov{MS}$):
\begin{equation}
  \alpha_1 = \frac{5}{3}\alpha_Y,\quad
  \alpha_Y = \frac{\alpha}{\cos^2\theta_{\ov{MS}}},\quad
  \alpha_2 = \frac{\alpha}{\sin^2\theta_{\ov{MS}}},\quad
  \alpha_3 \equiv \alpha_s = \frac {g^2_s}{4\pi},     \lb{81y}
\end{equation}
where $\alpha$ and $\alpha_s$ are the electromagnetic and SU(3)
fine structure constants respectively, $Y$ is the hypercharge, and
$\theta_{\ov{MS}}$ is the Weinberg weak angle in $\ov{MS}$ scheme.
Using RGEs with experimentally established parameters,
it is possible to extrapolate the experimental
values of three inverse running constants $\alpha_i^{-1}(\mu)$
(here i=1,2,3 correspond to U(1), SU(2) and SU(3) groups of SM)
from the Electroweak scale to the Planck scale.

It is well known (see for example \ct{25p})
that (in the absence of monopoles) the one--loop
approximation RGEs can be described by the following expressions:
\begin{equation}
  \alpha_i^{-1}(\mu) =
  \alpha_i^{-1}(\mu_R) + \frac{b_i}{4\pi}t
                                                \lb{4z}
\end{equation}
with slopes $b_i$ given by the following values:
$$
   b_i = (b_1, b_2, b_3) =
$$
\begin{equation}
( - \frac{4N_{gen}}{3} -\frac{1}{10}N_S, \quad
      \frac{22}{3}N_V - \frac{4N_{gen}}{3} -\frac{1}{6}N_S, \quad
      11 N_V - \frac{4N_{gen}}{3} ).                   \lb{5z}
\end{equation}
The integers $N_{gen},\,N_S,\,N_V\,$ are respectively the numbers
of generations, Higgs bosons and different vector gauge fields.

In SM we have:
\begin{equation}
       N_{gen} = 3, \quad N_S = N_V =1,                    \lb{6z}
\end{equation}
and the corresponding slopes (\ref{5z}) describe the evolutions of
$\alpha_i^{-1}(\mu)$.

   The precision of the LEP data allows to make the extrapolation of RGEs
with small errors up to the Planck scale (see \ct{33a}) unless the
new physics pops, of course. Assuming that these
RGEs for $\alpha_i^{-1}(\mu)$ contain only the contributions of the SM
particles up to $\mu=\mu_{Pl}$ and doing the extrapolation with one Higgs
doublet under the assumption of a "desert" and absence of monopoles,
we have the following result obtained in Ref.\ct{33a}:
\begin{equation}
   \alpha_1^{-1}(\mu_{Pl})\approx 33.3; \quad
   \alpha_2^{-1}(\mu_{Pl})\approx 49.5; \quad
   \alpha_3^{-1}(\mu_{Pl})\approx 54.0.
                                                        \lb{82y}
\end{equation}
The extrapolation of $\alpha_{1,2,3}^{-1}(\mu)$ up to the point
$\mu=\mu_{Pl}$ is shown in Fig.1 as function of the variable
$x=log_{10}\mu$ (GeV).

In this connection, it is very attractive to consider also the gravitational
interaction.

\subsection{"Gravitational fine structure constant" evolution}

The gravitational interaction between two particles
of equal masses M is given by the usual classical Newtonian potential:
\begin{equation}
   V_g = - G \frac{M^2}{r} =
           - \left(\frac{M}{M_{Pl}}\right)^2\frac{1}{r}
                   = - \frac{\alpha_g(M)}{r},              \lb{1x}
\end{equation}
which always can be imagined as a tree--level approximation of quantum
gravity.

Then the quantity:
\begin{equation}
      \alpha_g = \left(\frac{\mu}{\mu_{Pl}}\right)^2     \lb{2x}
\end{equation}
plays a role of the running "gravitational fine structure constant"
and the evolution of its inverse quantity also is presented
in Fig.1 together with the evolutions of $\alpha_i^{-1}(\mu)$.

Then we see the intersection of $\alpha_g^{-1}(\mu)$
with $\alpha_1^{-1}(\mu)$ at the point:
$$
               (x_0, \alpha_0^{-1}),
$$
where ( $\,x_0 = \log_{10}\mu_{int.}$):
\begin{equation}
      x_0 \approx 18.3,  \quad
       \alpha_0^{-1} \approx 34.4.                   \lb{3x}
\end{equation}

\section{Dropping of the Monopole Charge in the Family Replicated Gauge Group
Model (FRGGM)}

In the simplest case, the scalar monopole beta--function in QEMD is
is the same one as for the scalar QED (see \ct{20s},\ct{52a} and \ct{51}):
\be
   \beta(\tilde \alpha) = \frac{\tilde \alpha}{12\pi } +
{(\frac{\tilde \alpha}{4\pi })}^2 + ... = \frac{\tilde \alpha}{12\pi }( 1 +
 3\frac{\tilde \alpha}{4\pi } + ...).                \lb{1d}
\ee
It follows from Eq.(\ref{1d}) that the theory of monopoles cannot be considered
perturbatively  at least for
\be
          \tilde \alpha > \frac{4\pi}{3}\approx 4.       \lb{2d}
\ee
This limit is smaller for non--Abelian monopoles.

Using the Dirac relation (\ref{7}), it is easy to estimate in the simple SM
the Planck scale value $\tilde \alpha(\mu_{Pl})$
(minimal for $U(1)_Y$ gauge group):
\be
        \tilde \alpha(\mu_{Pl}) = \frac{5}{3}\alpha_1^{-1}(\mu_{Pl})/4
           \approx 55.5/4 \approx 14.                   \lb{3d}
\ee
This value is really very big compared with the estimation
which follows from (\ref{2d}) and, of
course, with the critical coupling $\tilde \alpha_{crit}\approx1$,
corresponding to the confinement---deconfinement phase
transition in the $U(1)_Y$ lattice gauge theory (see below Section
6). Clearly we cannot do the perturbation approximation with such
a strong coupling $\tilde{\alpha}$. It is hard for such monopoles
not to be confined. If they are confined
--- most easily by having an electrically charged scalar field condensate
--- they are effectively out of the game for scales much
smaller in energy $\mu$ than the mass $\mu_h=\mu_{"hadron"}$ of their "hadron
particles" --- bound states of them. With the huge $\tilde \alpha$
practically no running brings them out of the range of being huge and so
really the $\mu_h$ scale should go (at least) to the cut off scale.
This really simply means that such monopoles do not exist at all.
So if we make the formal assumption that monopoles exist and combine it
with the above suggestion, we get as conclusion that so strong
$\tilde \alpha $ is impossible.

There is an interesting way out of this problem if one wants to
have the existence of monopoles, namely to extend the SM gauge
group and let the monopoles correspond to charges which are really
linear combinations of the SM and extended ones. Then we might
attempt to choose a model so cleverly that certain selected linear
combinations get bigger electric couplings than the corresponding
SM couplings. That could make the monopoles which for these
certain charge linear combinations couple more weakly and thus
have a better chance of being allowed "to exist", in agreement
with the arguments for the confinement of too strongly coupled
particles.

An example of such an extension of SM that can impose the
possibility of the allowance of monopoles is just Family
Replicated Gauge Group Model (FRGGM).

\subsection{Family Replicated Gauge Group Model}

Most efforts to explain the Standard Model (SM) describing well the
experimental results known today are devoted to Grand Unification
Theories (GUTs). The supersymmetric extension of the SM consists of taking the
SM and adding the corresponding supersymmetric partners. The Minimal
Supersymmetric Standard Model (MSSM) shows \ct{33a} the possibility of the
existence of the grand unification point at
\begin{center}
$\mu_{GUT}\sim 10^{16}$ GeV.
\end{center}
Unfortunately, at present time experiment does not indicate any
manifestation of supersymmetry. In this connection, the
Anti--Grand Unification Theory (AntiGUT) was developed in
Refs.\ct{2a}, \ct{2e}-\ct{2k}, \ct{35}-\ct{38}, \ct{20p}-\ct{24p},
\ct{17p}-\ct{37} as a realistic alternative to SUSY GUTs.
According to this theory, supersymmetry does not come into the
existence up to the Planck energy scale (\ref{2}), where the
multiple critical point \ct{17p} exists and governs the behaviour
of the running coupling constants.

The Standard Model (SM) is based on the group SMG described by Eq.(\ref{1}). AntiGUT
suggests that at the scale $\mu_G\sim \mu_{Pl}=M_{Pl}$
there exists the more fundamental symmetry described by the family
replicated gauge group $G$ containing $N_{fam}$ copies
of the Standard Model Group SMG:
\begin{equation}
G = SMG_1\times SMG_2\times...\times SMG_{N_{fam}}\equiv (SMG)^{N_{fam}},
                                                  \lb{76y}
\end{equation}
where $N_{fam}$ designates the number of families.

If $N=3$ (as AntiGUT predicts~\cite{2k} and experiment confirms),
then the fundamental gauge group G is:
\begin{equation}
    G = (SMG)^3 = SMG_{1st\;fam.}\times SMG_{2nd\;fam.}\times SMG_{3rd\;fam.},
                                        \lb{77y}
\end{equation}
or the generalized one:
\begin{equation}
         G_f = (SMG)^3\times U(1)_f,           \lb{78y}
\end{equation}
which was suggested by the fitting of fermion masses of the SM
(see Refs.\ct{35}).

Recently a new generalization of AntiGUT was suggested in
Refs.\ct{38a},\ct{38}:
\begin{equation}
           G_{\mbox{ext}} = (SMG\times U(1)_{(B-L)})^3  \\
\equiv [SU(3)_c]^3\times [SU(2)_L]^3\times [U(1)_Y]^3\times
[U(1)_{(B-L)}]^3,    \lb{79y}
\end{equation}
which takes into account the see--saw mechanism with right-handed
neutrinos, describes all modern neutrino experiments, and gives
the reasonable fitting of the SM fermion masses and mixing angles.

Sometimes, for simplicity, we call the theory with the family
replicated gauge group of symmetry $G$, or $G_f$, or $G_{ext}$,
given by Eqs.(\ref{76y})-(\ref{79y}), "G--theory", especially
prefering to use this terminology for figures (see for example
Fig.2).

The group $G=G_{ext}$ contains: $3\times 8 = 24$ gluons, $3\times
3 = 9$ W-bosons and $3\times 1 + 3\times 1 = 6$ Abelian gauge
bosons.

At first sight, this ${(SMG\times U(1)_{(B-L)})}^3$ group with its 39
generators seems to be just one among many possible SM gauge group
extensions. However, it is not such an arbitrary choice. There are
at least reasonable requirements (postulates) on the gauge group G
which have uniquely to specify this group. It should obey the
following postulates (the first two are also valid for SU(5) GUT):

\vspace{0.1cm}

1. $G=G_{ext}$ should only contain transformations, transforming
the well--known 45 Weyl fermions of the SM ( = 3 generations of 15
Weyl particles each --- counted as left handed), and the
additional three heavy see--saw (right handed ) neutrinos --- into
each other unitarily, so that $G=G_{ext}$ must be a subgroup of
U(48): $G\subseteq U(48)$.

\vspace{0.1cm}

2. No anomalies, neither gauge nor mixed. Otherwise the model becomes
non--renormalisable. G--theory assumes that only
straightforward anomaly cancellation takes place and forbids the
Green-Schwarz type anomaly cancellation \ct{39}.

\vspace{0.1cm}

3. G--theory should NOT UNIFY the irreducible representations under the SM
gauge group, called here SMG (see Eq.(\ref{1})).

\vspace{0.1cm}

4. G is the maximal group satisfying the above-mentioned postulates.

\vspace{0.1cm}

In Refs.\ct{38a},\ct{38} the gauge group $G=(SMG \times
U(1)_{(B-L)})^3$ is spontaneously broken down (at one or two orders
of magnitude below the Planck scale) by 7 different Higgs fields
to the gauge group $SMG\times U(1)_{(B-L)}$ which is the diagonal
subgroup of G. Therefore, these Higgs fields break AGUT to the SM.
The field $\phi_{WS}$ corresponds to the Weinberg--Salam theory
(its VEV is known: $<\phi_{WS}>=246$ GeV), so that we have only
six free parameters -- six VEVs -- to fit the experiment in the
framework of this model. The authors of Refs.\ct{38} used them
with aim to find the best fit to conventional experimental data
for all fermion masses and mixing angles in the SM, also with aim
to explain the experiments in neutrino oscillations. The typical
fit to the masses and mixing angles for the SM leptons and quarks
is encouraging in the crude approximation. It was shown that the
family replicated gauge group $G_{ext}$ generates the large mixing
angle solution to the solar neutrino oscillation problem. Also the
neutrino masses are predicted in Ref.\ct{38}.

Finally, we conclude that, in general, the theory with
$FRGG$--symmetry is very successful in describing of the SM
experiment.

\subsection{FRGG-model prediction for the values of electric and magnetic
charges}

According to the FRGG--model, at some point $\mu=\mu_G < \mu_{Pl}$
(or really in a couple of steps)
the fundamental group $G$ undergoes spontaneous breakdown to its
diagonal subgroup:
\begin{equation}
      G \longrightarrow G_{diag.subgr.} = \{g,g,g || g\in SMG\},
                                                          \lb{83y}
\end{equation}
which is identified with the usual (low--energy) group SMG.
The point $\mu_G\sim 10^{18}$ GeV is shown in Fig.2 demonstrating the
evolution of the inverses $\alpha_{Y,2,3}^{-1}$ up to the Planck scale.
But it seems that this extrapolation is valid only up to the $\mu = \mu_G$,
because FRGGM (here named G-theory) works in the region
$\mu_G \le \mu \le \mu_{Pl}$, giving the quite different behaviour of
$\alpha_i^{-1}(\mu)$ in this region, which is not shown in Fig.2, but
will be investigated in Section 9.

It should be said that in the FRGG--model
each family has its own gluons, own W's, etc. The breaking just
makes only a certain linear combination of gluons which
exists in the SM: below $\mu=\mu_G$ and down to the low energies.
We can say that the phenomenological gluon is a linear
combination (with amplitude $1/\sqrt 3$ for $N_{fam}=3$) for each of the
FRGG gluons of the same colour combination.
Then we have the following formula connecting the fine structure constants
of non--Abelian FRGG--model to the low energy surviving diagonal subgroup
$G_{\mbox{diag.subg.}}\subseteq {(SMG)}^3$:
\begin{equation}
\alpha_{i,\mbox{diag}}^{-1} = \alpha_{i,\mbox{1st\, fam.}}^ {-1} +
\alpha_{i,\mbox{2nd\, fam.}}^{-1} + \alpha_{i,\mbox{3rd\,
fam.}}^{-1}.
                                                      \lb{86yA}
\end{equation}
Here i = SU(2), SU(3), and i=3 means that we talk about the gluon
couplings.

Assuming that three FRGG couplings are equal to each other
(as comes indeed out of our MPP, see later the Section 5), we obtain:
\begin{equation}
\alpha_{i,\mbox{diag}}^{-1}\approx 3\alpha_{i,\mbox{one fam.}}^
{-1} \equiv 3\alpha_{i,G}^{-1}.
                                                      \lb{86yB}
\end{equation}
In contrast to non-Abelian theories, in which the gauge invariance
forbids the mixed (in families) terms in the Lagrangian of
FRGG--theory, the U(1)--sector of FRGG contains such mixed
terms (see Ref.\ct{17p}):
\begin{equation}
\frac{1}{g^2}\sum_{p,q} F_{\mu\nu,\; p}F_{q}^{\mu\nu} =
\frac{1}{g^2_{11}}F_{\mu\nu,\; 1}F_{1}^{\mu\nu} +
\frac{1}{g^2_{12}}F_{\mu\nu,\; 1}F_{2}^{\mu\nu} +
...
+ \frac{1}{g^2_{23}}F_{\mu\nu,\; 2}F_{3}^{\mu\nu} +
\frac{1}{g^2_{33}}F_{\mu\nu,\; 3}F_{3}^{\mu\nu},
                                                            \lb{87y}
\end{equation}
where $p,q = 1,2,3$ are the indices of three families of the group
$(SMG)^3$. Now it is easily seen that if the different families
had specific equal electric charges, i.e. equal $\alpha_{pq}$,
then taking the diagonal subgroup we obtain \ct{17p}:
\begin{equation}
\alpha_{\mbox{diag}}^{-1}\approx 6\alpha_{\mbox{G}}^ {-1},
                                                      \lb{87yB}
\end{equation}
which shows that we can increase electric $\alpha$ by a factor 6
replacing it by the electric $\alpha_{\mbox{one\, fam.}}\equiv
\alpha_G$.

In the article \ct{17p} it was estimated that a large number (presumably
largest possible) phases might be brought to meet by having a certain
hexagonal symmetry which actually would mean that various "cross couplings"
$g_{pq}$ in (\ref{87y}) are equal to each other.

Taking (\ref{87yB}), we can get the new $\tilde \alpha_{one\, fam.}$
for the U(1)--sector of FRGG which is
smaller by factor 6 in comparison with $\tilde \alpha$ in SM:
\begin{equation}
             \tilde \alpha_G = \frac {1}{6} \tilde \alpha_{diag}.  \lb{4da}
\end{equation}
Using the result (\ref{3d}), we can estimate $\tilde \alpha $
for one family at the Planck scale:
\begin{equation}
        \tilde \alpha_G(\mu_{Pl}) \approx 14/6 \approx 2.3 .
                                                           \lb{4d}
\end{equation}
But it seems that $\tilde \alpha_{G}(\mu_{Pl}) \approx 1$
in FRGGM (see Section 8), and the
perturbation theory works for $\beta$--function of scalar
monopoles near the Planck scale.

In general, considering the non--Abelian FRG--group $SU(N)^{N_{fam}}$ we
have the following value for the one family monopole charge:
\begin{equation}
                  \tilde \alpha_G = \frac{\tilde \alpha}{N_{fam}},
\lb{4y}
\end{equation}
where $\tilde \alpha$ is the total fine structure constant.

But for $U(1)^{N_{fam}}$ gauge group, taking into account the
considerations
of Ref.\ct{17p}, it is easy to obtain the following relation:
\begin{equation}
     \tilde \alpha_{1,G} = \frac{\tilde \alpha_1}{N^*} \quad,
      {\mbox{where}}  \quad N^* = \frac{1}{2}N_{fam}(N_{fam}+1).   \lb{4z}
\end{equation}
The conclusion: if one wants monopoles "to exist", it is necessary to
drive in the direction of a model like FRGG.

\section{Multiple Point Principle}

AntiGUT approach is often used in conjunction with the Multiple
Point Principle (MPP) proposed by D.L.Bennett and H.B.Nielsen
\ct{17p}. According to this principle, Nature seeks a special
point -- the Multiple Critical Point (MCP) -- which is a point on
the phase diagram of the fundamental regularized gauge theory G,
where the vacua of all fields existing in Nature are degenerate
having the same vacuum energy density. Such a phase diagram has
axes given by all coupling constants considered in theory. Then
all (or just many) numbers of the phases meet at the MCP.

MPM assumes the existence of MCP at the Planck scale,
insofar as gravity may be "critical" at the Planck scale.
\footnote{ We say "at the Planck scale" because we have in mind the definition
of the concept of phase transition representing the borderline in space
of couplings cross producted with the axis of scales so that we think
of phase transitions being at different coupling values depending on
the scales considered. There is another, we think more usual, way of thinking
of it: a phase transition is along a hypersurface in the space of only
parameters of the theory, such as couplings, the temperature and the mass
parameters, where the various output properties have jumps.}

The philosophy of MPM leads to the need for
investigating the phase transition in various gauge theories.
A lattice model of gauge theories is the most convenient formalism
for the realization of the MPM ideas. As it was mentioned above,
in the simplest case we can imagine our
space--time as a regular hypercubic (3+1)--lattice with the parameter $a$
equal to the fundamental (Planck) scale:
\begin{equation}
    a = \lambda_P = 1/M_{Pl}\sim 10^{-33}\,{\mbox {cm}}.    \lb{18a}
\end{equation}
In general, the lattice results are very encouraging for MPM.
Taking seriously the ideas of Ref.\ct{17p}, we were able to predict gauge
coupling constants in terms of lattice phase transition couplings, using very
strongly FRGG--model.

It is necessary to emphasize, as it was mentioned in the introduction,
that we do not know, what lattice--like structure plays role in
the description of physical processes at very small distances. We
assume and have tried to show (see the review \ct{11}) that one
can crudely calculate the phase transition couplings without using
any specific lattice, rather only approximating the lattice
artifact monopoles as fundamental (pointlike) particles
condensing. The details of the lattice --- hypercubic or random,
with multiplaquette terms or without them, etc. --- also the
details of the regularization --- lattice or Wilson loops, lattice
or Higgs monopole model --- do not matter too much for the value
of the phase transition couplings. Critical couplings depend only
on the group, not so much on the regularization. Such an approximate
universality of critical couplings is, of course, absolutely
needed if there should be any sense in relating lattice phase transition
couplings to the experimental couplings found in Nature.
Otherwise, such a comparison would only make sense if we could
guess the true lattice in the right model, what sounds too
ambitious.

\section{Phase Transitions in Lattice Gauge Theories}

Following the MPM ideas, we consider our space--time as a simple regular
hypercubic (3+1)--lattice with the parameter $a$ equal to the fundamental
Planck scale (\ref{18a}). Our aim is to investigate the phase transitions
in the lattice gauge theories and relate the obtained phase transition
couplings with MCP at the Planck scale. The lattice investigations were
performed by a lot of authors \ct{1sa}-\ct{13p} using Monte Carlo method
of simulations of the lattice gauge theories, and we have considered
the Higgs monopole model \ct{20p}-\ct{24p} to explain their results.

\subsection{Lattice actions}

A lattice contains sites, links and plaquettes.
Link variables are fundamental
variables of the lattice theory. These variables are also
the elements of gauge group $G$, describing a symmetry of the
corresponding lattice gauge theory:
\begin{equation}
{\cal U}(x\link y) \in G.
\lb{18}
\end{equation}
The link variable connects the point $n$ and the point
$n + a_{\mu}$, where the index $\mu$ indicates the direction of a link
in the hypercubic lattice with parameter $a$:
\begin{equation}
   {\cal U}(x\link y) = e^{i\Theta_{\mu}(n)} \equiv {\cal U}_{\mu}(n),
                                                                 \lb{22}
\end{equation}
where
$$
\Theta_{\mu}(n) = a{\hat A}_{\mu}(x),
$$
$$
  {\hat A}_{\mu}(x) = gA_{\mu}(x)\quad{\mbox{--- for U(1)-group,}}
$$
\be
 {\hat A}_{\mu}(x) = gA^j_{\mu}(x) t^j\quad{\mbox{--- for SU(N)-group}}.
                                                       \lb{24}
\ee
Here $t^j$ is the generator of the SU(N)-group, and
$\,t^j = \lambda^j/2$ for SU(3)-group,
where $\lambda^j$ are the well--known Gell--Mann matrices.

Plaquette variables are products of link variables:
\begin{equation}
{\cal U}_p\equiv {\cal U}(\square){\stackrel{def}{=}}
      {\cal U}\mbox{\plaqb} {\cal U}\mbox{\plaqr} {\cal U}\mbox{\plaqt}
      {\cal U}\mbox{\plaql}.
                                             \lb{26}
\end{equation}
The simplest action of the U(1) lattice gauge theory is given by the
following expression:
\begin{equation}
     S[{\cal U}_p] = \beta \sum_p \cos {\Theta}_p,\quad
{\mbox{where}} \quad {\cal U}_p = e^{i\Theta_p}.              \lb{35a}
\end{equation}
For the compact lattice U(1) theory: $\beta=1/e_0^2$, where $e_0$ is the
bare charge.

The Villain form of the U(1) lattice gauge theory \ct{1sa} is:
\begin{equation}
   S_V = (\beta/2) \sum_p {({\Theta}_p - 2\pi k_p)}^2, \quad k_p\in Z,
                                          \lb{35b}
\end{equation}
where $k_p$ is defined to be that integer giving the term of the smallest
value.

The lattice $SU(N)$ gauge theories were first introduced by K.Wilson
\ct{1s} for studying the problem of confinement. He suggested the following
simplest action:
\begin{equation}
S = - \frac{\beta}{N}\sum_p Re(Tr\,{\cal U}_p),          \lb{36}
\end{equation}
where the sum runs over all plaquettes of a hypercubic lattice.

Monte Carlo simulations of these simple Wilson lattice
theories in four dimensions showed a (or an almost) second--order
deconfining phase transition for U(1) \ct{2s}, a crossover
behaviour for SU(2) and SU(3) \ct{4s}, and a first--order
phase transition for SU(N) with $N\ge 4$ \ct{6s}.

Bhanot and Creutz \ct{7s} have generalized the simple Wilson theory,
introducing two parameters in the $SU(N)$ action:
\begin{equation}
   S = \sum_p[-\frac{\beta_f}{N}Re(Tr\,{\cal U}_p) -
               \frac{\beta_A}{N^2-1}Re(Tr_A{\cal U}_p)],   \lb{37}
\end{equation}
where $\beta_f$, $Tr$ and $\beta_A$, $Tr_A$ are
respectively the lattice constants and traces in the fundamental and
adjoint representations of $SU(N)$.

Monte Carlo simulations of the U(1) gauge theory described by the
two-parameter lattice action \ct{9s},\ct{10s}:
\begin{equation}
     S = \sum_p[\beta^{lat} \cos \Theta_p + \gamma^{lat} \cos2\Theta_p]
                                              \lb{38}
\end{equation}
show the existence of a triple point on the corresponding
phase diagram: "Coulomb--like", totally confining and $Z_2$ confining
phases come together at this triple point.

The phase diagrams were obtained for the generalized lattice SU(2) and
SU(3) theories (\ref{37}) by Monte Carlo methods in Refs.\ct{7s}
(see also \ct{8sa}). They indicated
the existence of a triple point which is a boundary point of three
first--order phase transitions: the "Coulomb--like" and the confining
phase and that phase in which $Z_N$ subgroup confines, while the factor
group $SU(N)/Z_N$ does not confine, make up the three
phases that meet at this point. From the triple
point emanate three phase border lines which separate the corresponding
phases. All confinement phases exist due to monopole condensations of
slightly different sorts.

\subsection{The behaviour of the lattice electric fine structure constant
$\alpha$ near the phase transition point}

The lattice investigations did not give the lattice
triple point values of $\alpha_{i,crit}$ by Monte Carlo simulation
method. Only in Ref.\ct{10s} the critical value of the effective electric
fine structure constant $\alpha$ was obtained in the compact U(1)
lattice gauge theory described
by the Wilson and Villain actions (\ref{35a}) and (\ref{35b}), respectively:
\begin{equation}
\alpha_{crit}^{lat}\approx 0.20\pm 0.015\quad
{\mbox{and}} \quad {\tilde \alpha}_{crit}^{lat}\approx 1.25\pm 0.10
\quad{\mbox{at}}\quad
\beta_T\equiv\beta_{crit}\approx{1.011}.
                                                     \lb{47}
\end{equation}
The result of Ref.\ct{10s} for the behaviour of $\alpha(\beta)$ in the vicinity
of the phase transition point $\beta_T$ is shown in Fig.3 for the Wilson
and Villain lattice actions. Fig.4 demonstrates the comparison of the
function $\alpha^{-1}(\beta)$ obtained by Monte Carlo method for the Wilson
lattice action and by our theoretical calculation of the same quantity.
The theoretical (dashed) curve was calculated by so-called "Parisi improvement
formula" \ct{13p}:
\begin{equation}
    \alpha (\beta )=[4\pi \beta W_p]^{-1}.     \lb{48}
\end{equation}
Here $W_p=<\cos \Theta_p >$ is a mean value of the plaquette energy.
The corresponding values of $W_p$ were taken from Ref.\ct{9s}.

The theoretical value of $\alpha_{crit}$ is less than the "experimental"
(Monte Carlo) value (\ref{47}):
\begin{equation}
      \alpha_{crit.,theor.}\mbox{(in\,\,lattice\,\,theory)}\approx{0.12}.
                                                    \lb{49}
\end{equation}
According to Fig.4:
\begin{equation}
   \alpha_{crit.,theor.}^{-1}\approx 8,
\quad{\mbox{and}}\quad
   \alpha_{crit.,lat.}^{-1}\approx 5.                   \lb{50a}
\end{equation}
This discrepancy between the theoretical and "experimental"
results, shown in Fig.4, has the following explanation: "Parisi
improvement formula" (\ref{48}) does not take into account the
contribution of monopoles running due to the renormalisation group
near the phase transition point.

\subsection{Lattice artifact monopoles and Higgs Monopole Model}

Lattice monopoles are responsible for the confinement in lattice gauge
theories what is confirmed by many numerical and theoretical investigations
(see reviews \ct{11s} and papers \ct{12s}).
In the compact lattice gauge theory the monopoles are not physical objects:
they are lattice artifacts driven to infinite mass in the continuum
limit.

In the papers \ct{17p}-\ct{19p} the calculations of the U(1)
phase transition (critical) coupling constant were connected with the
existence of artifact monopoles in the lattice gauge theory and also
in the Wilson loop action model \ct{19p}.

In Ref.\ct{19p} we have put forward the speculations of
Refs.\ct{17p} and \ct{18p} suggesting that the modifications of the
form of the lattice action might not change too much the phase
transition value of the effective continuum coupling constant. The
purpose was to investigate this approximate stability of the
critical coupling with respect to a somewhat new regularization
being used instead of the lattice, rather than just modifying the
lattice in various ways. In \ct{19p} the Wilson loop action was
considered in the approximation of circular loops of radii $R\ge
a$. It was shown that the phase transition coupling constant is
indeed approximately independent of the regularization method:
${\alpha}_{crit}\approx{0.204}$, in correspondence with the Monte Carlo
simulation result on lattice \ct{10s} given by
Eq.(\ref{47}). It was the first step to confirm the "approximate
universality" of the critical couplings which we have considered
in other our papers (see review \ct{11}).

In Refs.\ct{20p}-\ct{24p}, instead of using the lattice or Wilson loop
cut--off, we have developed the Higgs Monopole Model (HMM) approximating
the lattice artifact monopoles as fundamental pointlike particles described
by the Higgs scalar field.
Indeed, the simplest effective dynamics describing the
confinement mechanism in the pure gauge lattice U(1) theory
is the dual Abelian Higgs model of scalar monopoles \ct{11s},\ct{12s}.
This model has the following Lagrangian:
\begin{equation}
    L = - \frac{1}{4g^2} F_{\mu\nu}^2(B) + \frac{1}{2} |(\partial_{\mu} -
           iB_{\mu})\Phi|^2 - U(\Phi),              \lb{5y}
\end{equation}
where
\begin{equation}
 U(\Phi) = \frac{1}{2}\mu^2 {|\Phi|}^2 + \frac{\lambda}{4}{|\Phi|}^4
                                       \lb{6y}
\end{equation}
is the Higgs potential of scalar monopoles with magnetic charge $g$, and
$B_{\mu}$ is the dual gauge (photon) field interacting with the scalar
monopole field $\Phi$.  In this model the positive $\lambda$ is the
self--interaction constant of scalar fields, whereas the mass parameter
$\mu^2$ is negative.

Considering the renormalisation group improvement
of the effective Coleman--Weinberg potential \ct{20s} (see also \ct{21s}),
written in Refs.\ct{22p}-\ct{24p} for the dual sector of scalar
electrodynamics in the two--loop approximation for
$\beta$--functions, we have calculated the U(1)
critical values of the magnetic fine structure constant:
\begin{equation}
       {\tilde\alpha}_{crit} = g^2_{crit}/4\pi\approx 1.20    \lb{51A}
\end{equation}
and (by the Dirac relation) electric fine structure constant:
\begin{equation}
     \alpha_{crit} = \pi/g^2_{crit}\approx 0.208.               \lb{51B}
\end{equation}
These values coincide with the lattice result (\ref{47}).

Here it is quite necessary to comment that the lattice result
$\alpha_{crit}\approx 0.2$ is valid for U(1) lattice gauge theory with or
without electrically
charged fermions (at least approximately), what is easy to confirm
calculating the partition function with fermions on lattice and
integrating over fermion fields \ct{52}.
By this reason, considering the U(1) theory with fermions
in the vicinity of the phase transition point, we have
used the lattice results for pure U(1), what
is very important for the predictions of MPM (see Section 8).

\section{ The Evolution of the Running U(1) fine structure Constant
in the Theory with/without the Higgs Monopole Field}

Considering the evolution of the running U(1) fine
structure constant $\alpha_Y(\mu)=\alpha(\mu)/cos^2\theta_{\overline{MS}}$,
we have in SM the following one--loop approximation:
\begin{equation}
  \alpha_Y^{-1}(\mu) =
  \alpha_Y^{-1}(\mu_R) + \frac{b_Y}{4\pi}t,
                                                \lb{4x}
\end{equation}
where $t$ is the evolution variable (\ref{6}), and $b_Y$ is given
for $N_{gen}$ generations and $N_S$ Higgs bosons by the
following expression:
\begin{equation}
        b_Y = - \frac{20}{9}N_{gen} - \frac{1}{6} N_S.
                                                     \lb{4xb}
\end{equation}
The evolution of $\alpha_Y^{-1}(\mu)$ is represented in Fig.2 for
$N_{gen}=3$ and $N_S=1$ by the straight line going up to $\mu = \mu_{Pl}$.

Let us consider now the exotic (not existing in reality) case when
we have, for example, the cut off energy $\mu_{cut off}\sim 10^{42}$ GeV.
In this case the evolution of ${\alpha_Y}^{-1}(\mu)$ is given by Fig.5a,
where the straight line 1 (one--loop approximation) goes to the Landau pole
at $\alpha_Y^{-1} = 0$. But it is obvious that in the vicinity of the Landau
pole, when $\alpha_Y^{-1}(\mu)\to 0$, the charge of U(1) group becomes larger
and larger with increasing of $\mu$. This means that the one--loop
approximation for $\beta$--function is not valid for large $\mu$,
and the straight line 1 may change its behaviour.
In general, the two--loop approximation for $\beta$--function of QED
(see Refs.\ct{51},\ct{20s} and \ct{52a}) shows that this straight
line turns and goes down.

Exotically we can consider our space-time as a lattice with parameter $a$
smaller than the Planck scale value $\lambda_P$. For example,
we can imagine a lattice with $a\sim (10^{42}\,{\mbox{GeV}})^{-1}$,
or the existence of the fundamental magnetically charged Higgs scalar field
in the vicinity of large $\mu_{crit}\sim 10^{38}$ GeV, when we have the
phase transition point with $\alpha_{Y,crit}^{-1}\approx 4$ (see below
(\ref{5x}) and Fig.5a).

The artifact monopoles, responsible for the confinement of electric charges
at the very small distances, can be approximated
by the magnetically charged Higgs scalar field, which
leads to the confinement--deconfinement phase transition, as it was shown
in Refs.\ct{20p}-\ct{24p}. If we have this phase transition, then there
exists a rapid fall in the evolution of $\alpha^{-1}(\mu)\,\,
(\alpha_Y^{-1}$ or $\alpha_{Y,G}^{-1}$, etc.) near the phase transition
(critical) point. This "fall" always accompanies the phase transition from
the Coulomb--like phase to the confinement one.

Indeed, we can present the effective Lagrangian of our field system as a
function of the variable
\begin{equation}
F^2\equiv F_{\mu\nu}^2,     \lb{5xd}
\end{equation}
where
\begin{equation}
 F_{\mu\nu}= \partial_{\mu}A_{\nu}-\partial_{\nu}A_{\mu}   \lb{6x}
\end{equation}
is the field strength tensor:
\begin{equation}
L_{eff}  =
       - \frac {\alpha_{eff}^{-1}(F^2)}{16\pi} F^2.
                                   \lb{6xab}
\end{equation}
When $F^2={\vec B}^2$ ($\vec B$ is the magnetic field)
and ${\vec B}^2$ is independent of
space--time coordinates (see Refs.\ct{22q},\ct{23q}), we
can write the effective potential:
\begin{equation}
V_{eff}  =
         \frac {\alpha_{eff}^{-1}({\vec B}^2)}{16\pi} {\vec B}^2 =
           A\alpha_{eff}^{-1}(t)e^{2t}.
                                               \lb{6xa}
\end{equation}
Here, choosing ${\vec B}^2={\mu}^4$ and $\mu_R=\mu_{cutoff}$, we have:
$$
A={\mu_{cutoff}^4}/{16\pi},
$$
and
\begin{equation}
            t = \frac{1}{2}\log{\frac{F^2}{\mu_{cutoff}^4}}.
                                                      \lb{6xd}
\end{equation}
It is well--known \ct{22q},\ct{23q} that $\alpha_{eff}^{-1}(t)$ has the
same evolution over $t=\log (\mu^2/\mu_{cutoff}^2)$  whether we consider
$\mu^2=p^2$ (where p is the 4-momentum), or $\mu^4=F^2={\vec B}^2$
(at least, up to the second order perturbation).

In the confinement region ($t > t_{crit}$) the effective potential
(\ref{6xa}) has a minimum, given by the requirement \ct{23q}:
\begin{equation}
         [\frac {d\alpha_{eff}^{-1}}{dt} +
                           2\alpha_{eff}^{-1}(t)]|_{t=t_{min}} = 0.
                                     \lb{7x}
\end{equation}
Of course, we need this minimum for $t$--values above $t_{crit}$ in
order to have confinement which namely means that we have a nonzero $F^2
={\vec B}_0^2=const$ in the vacuum \ct{14p}-\ct{24q}.

This minimum of the effective potential can exist only if
$\alpha_{eff}^{-1}(\mu)$ has a rapid fall near the phase
transition point which is illustrated in Fig.5a,b by curve 2. Moreover, the
width of this fall must be narrow enough to give a minimum of $V_{eff}$.
Indeed, the condition $V'_{eff} < 0$ gives:
\begin{equation}
 [\frac{d\alpha_{eff}^{-1}}{dt} + 2\alpha_{eff}^{-1}(t)]|_{t=t_{crit}} < 0,
                                          \lb{8x}
\end{equation}
or
\begin{equation}
   \Delta t  < - \frac
{\Delta(\alpha_{eff}^{-1})}{2\alpha_{eff}^{-1}}|_{t=t_{crit}}.  \lb{9x}
\end{equation}
If the value $\Delta \alpha_{eff}^{-1}$ is not large in the critical point,
then $\Delta t$ is small, and we have a sharp jump of $V_{eff}$ (what means
the almost first order phase transition) and a sharp jump of
$\alpha_{eff}^{-1}$ from the one-loop approximation end point value
(we call it $\alpha_{crit.Coul.}$) to the critical one ($\alpha_{crit.conf.}$).
This case is realized in AntiGUT and will be discussed
later in Section 8.

The existence of minimum of the effective potential (\ref{6xa})
explains why the straight line 1 changes its behaviour and rapidly
falls: $\alpha_{eff}^{-1}(t)$ is multiplied by $\exp(2t)$
in Eq.(\ref{6xa}).

After this "fall" $\alpha_Y^{-1}(\mu)\,$ has a
crook and goes to the constant value, as it is shown in Fig.5a by solid
curve 2, demonstrating the phase transition from Coulomb--like phase to the
confinement one.

The next step is to give the explanation why $\alpha_{eff}^{-1}(t)$
is arrested when $t\to t_{cutoff}$.

The process of formation of strings in the confinement phase (considered in
Ref.\ct{21p}) leads to the "freezing" of $\alpha$: in the confinement phase
the effective electric fine structure constant is almost unchanged and
approaches its maximal value $\alpha=\alpha_{max}$ when ($\mu \to \infty$).
The authors of Ref.\ct{14p} predicted that the fine structure constant
$\alpha$ cannot be infinitely large, but has the following maximal value:
\begin{equation}
\alpha_{max} \approx \frac{\pi}{12}\approx 0.26,
                                                         \lb{51}
\end{equation}
due to the Casimir effect for strings. This viewpoint was developed
in spinor QED by authors of Ref.\ct{14pa}: the vacuum polarization
induced by thin "strings"---vortices of magnetic flux---leads to the
suggestion of an analogue of the "spaghetti vacuum" \ct{14pb} as a possible
mechanism for avoiding the divergences of perturbative QED. According to
Ref.\ct{14pa}, the non--perturbative sector of QED arrests the growth
of the effective $\alpha$ to infinity for small distances and confirms
the existence of $\alpha_{max}$.
This phenomenon was called "the freezing of $\alpha$".

Fig.3 demonstrates the tendency to freezing of $\alpha$ in the compact QED
for $\beta < \beta_T$ (i.e. for "bare" constant $e_0 > 1$, what means
$\alpha_0^{-1} < 4\pi \approx 12.56$).

Choosing the lattice result (\ref{47}), which almost coincides with
the HMM result (\ref{51B}), we have:
\begin{equation}
     \alpha_{Y}^{-1}(\mu_{crit}) \equiv \alpha_{U(1),crit}^{-1}\approx 5.
                                                            \lb{5x}
\end{equation}
Analogously, using (\ref{51}), we have the maximal value
for $\alpha_Y^{-1}$:
\begin{equation}
       \alpha_{Y,max}^{-1}
        \approx 1/0.26 \approx 3.8.              \lb{5xa}
\end{equation}
An interesting situation arises in the theory with FRGG--symmetry,
when it begins to work at $\mu=\mu_G (< \mu_{Pl})$. As it was shown
in Section 4, in the vicinity of the phase transition point the
U(1)--sector of FRGG has $\alpha_{Y,G}\equiv
\alpha_{Y,one\,fam.}$, which is 6 times larger than $\alpha_Y$.
Now the phase transition "deconfinement--confinement" occurs
at $\mu_{crit}=\mu_{Pl}$ (but not at $\mu_{crit}\sim 10^{38}$ GeV
as it was in the SM prolonged up to the scale $\mu_{cutoff}\sim 10^{42}$
GeV).
This case is not "exotic" more, and confirms the MPP idea \ct{17p}.

The evolution of the inverse one family fine structure constant
$\alpha_{Y,G}(\mu )^{-1}(x)$  is given in Fig.5b, which
demonstrates the existence of critical point at the Planck scale.

Here it is necessary to comment that we have given a qualitative
behaviour of the fall in Fig.5a,b believing in the confinement
existing in the $U(1)_Y$ theory. In general,
the existence of the confinement (see \ct{14pa}),
as well as the shape of the fall (wide or narrow), depends on the type
of theory considered. If the cutoff energy is not very high
($\mu_{cutoff} < \mu_{crit}$), or the lattice spacing $a$ is not too small,
then
there is no confinement region in such a
theory (for example, in the U(1) sector of the SM we have
$\mu_{cutoff}=\mu_{Pl}$ which is smaller than $\mu_{crit}$, given by
Fig.5a, and the confinement phase is not available).

According to MPP {\footnote{We call it MPP-II to distinguish this
definition from the version in which MPP-I is defined only as a
phase transition depending on bare couplings while no scale
involved}}, Nature has to have the phase transition point at the
Planck scale not only for the Abelian U(1) theory, but also for
non--Abelian theories.
This means that the effective potential of
SU(3) gauge theory has the second minimum at the Planck scale
(the first one corresponds to the low--energy hadron physics).
String states of this second confinement phase are not observed in Nature,
because the FRGG--theory approaches the confinement phase at the Planck scale,
but does not reach it.

\section{FRGGM Prediction of the Planck Scale Values of the
U(1), SU(2) and SU(3) Fine Structure Constants}

The FRGGM prediction of the values of $\alpha_i(\mu)$ at $\mu=\mu_{Pl}$
is based on the MPM assumption about the existence of phase
transition boundary point MCP at the Planck scale, and gives these values
in terms of the corresponding critical couplings $\alpha_{i,crit} $
\ct{2c}-\ct{2k},\ct{17p}:
\begin{equation}
            \alpha_i(\mu_{Pl}) = \frac {\alpha_{i,crit}}{N_{fam}}
                       = \frac{\alpha_{i,crit}}{3}
      \quad{\mbox{for}}\quad i=2,3\quad\mbox{ (also for $i>3$),}       \lb{84y}
\end{equation}
and
\begin{equation}
        \alpha_1(\mu_{Pl}) = \frac {\alpha_{1,crit}}{{\frac 12}N_{fam}
        (N_{fam} + 1)}
           = \frac{\alpha_{1,crit}}{6} \quad{\mbox{for}}\quad U(1).
                                      \lb{85y}
\end{equation}
A simple explanation of the relations (\ref{84y}) and (\ref{85y})
was given in Section 4 by Eqs.(\ref{86yA})-(\ref{87y}).

\subsection{Phase transition couplings in the regularized SU(N) gauge theories}

It was shown in a number of investigations (see for example \ct{11s},\ct{12s}
and references there), that the confinement in the SU(N) lattice gauge
theory effectively comes to the same U(1) formalism. The reason is the
Abelian dominance in their monopole vacuum: monopoles of the Yang--Mills
theory are the solutions of the U(1)--subgroups, arbitrary embedded into
the SU(N) group. After a partial gauge fixing --- Abelian projection by
't Hooft \ct{24q} --- SU(N) gauge theory is reduced to an Abelian
$U(1)^{N-1}$ theory with $N-1$ different types of Abelian monopoles.
Choosing the Abelian gauge for dual gluons, it is possible to describe
the confinement in the lattice SU(N) gauge theories by the analogous
dual Abelian Higgs model of scalar monopoles.

Using the Abelian gauge by 't Hooft and taking into account that
the direction in the Lie algebra of monopole fields are gauge
dependent, we have found in Ref.\ct{22p} an average over these directions
and obtained \un{the group dependence relation} between the phase transition
fine structure constants for the groups $U(1)$ and $SU(N)/Z_N$ :
\begin{equation}
\alpha_{N,crit}^{-1} =
           \frac{N}{2}\sqrt{\frac{N+1}{N-1}} \alpha_{U(1),crit}^{-1}.
                                             \lb{25z}
\end{equation}
We have calculated this relation using only the one--loop approximation
diagrams of non-Abelian theories.

According to Eq.(\ref{25z}), we have the following relations:
\begin{equation}
    \alpha_{U(1),crit}^{-1} : \alpha_{2,crit}^{-1} : \alpha_{3,crit}^{-1}
           = 1 : \sqrt{3} : 3/\sqrt{2},
                                            \lb{26z}
\end{equation}
which together with relations (\ref{84y}) and (\ref{85y}) give:
\begin{equation}
\begin{array}{l}
\alpha_1^{-1}(\mu_{Pl}) \approx
6\cdot \frac{3}{5}\alpha_{U(1),crit}^{-1}\approx 18,
\\
\alpha_2^{-1}(\mu_{Pl}) \approx
3\cdot \sqrt{3}\alpha_{U(1),crit}^{-1}\approx 26,
\\
\alpha_3^{-1}(\mu_{Pl}) \approx
3\cdot \frac{3}{\sqrt 2}\alpha_{U(1),crit}^{-1}\approx 32,
\end{array}
                                                       \lb{7z}
\end{equation}
where we have used the value (\ref{5x}).

Here we see that in FRGGM the magnetic fine structure constant of
Abelian monopoles approximately is equal to unity at the Planck scale:
\be
   \tilde \alpha_{U(1)}(\mu_{Pl}) = \tilde \alpha_{U(1),crit}
            = \frac{1}{4}\alpha_{U(1),crit}^{-1}
             \approx \frac{5}{4}\approx 1.25,                     \lb{8z}
\ee

and the corresponding beta-function can be considered perturbatively
(see Section 2, inequality (\ref{22x})).

One should tell here that monopoles only escape confinement above
$\mu_G$ scale because there is the Higgsing of the gauge group G down
to the diagonal subgroup. These Higgs fields must somehow at least
confine the family monopoles.

\subsection{Olive's monopoles}

The fact that we have one special monopole for each of the three
groups SU(3), SU(2) and U(1), which we have considered when calculated
the phase transition (critical) couplings (for the confinement due to
monopole condensation, either in the SM, or in FRGGM) is indeed not
consistent with quarks and leptons as phenomenologically found objects.
The point is that the SU(3) monopole we used, for instance, radiates a
flux corresponding to a path in the gauge group SU(3) from the unit
element to one of the non--trivial center elements. Such a monopole
gives rise to a phase factor $exp(2\pi i/3)$ when a quark encircles
its Dirac string. Therefore, it does not allow quarks.

What is allowed consistently with the SM representations is what
D.Olive \ct{40a} has proposed and we call "the Olive--monopole".
These monopoles have the magnetic charge {\em under all three
subgroups: SU(3), SU(2) and U(1)} of SMG. Their total
magnetic charge corresponds to the center element
$ (Ie^{i2\pi /3},\, -I,\, 2\pi )\in SU(3)\times SU(2)\times {\bf R}$
contained in the covering group $SU(3)\times SU(2)\times {\bf R}$ of
SMG, which according to the interpretation of Ref.\ct{40b},
has a meaning of the {\underline{Lie group}}, rather than just Lie algebra,
fitting the SM representations. That is, Olive--monopole
in the SM has at the same time three different monopolic charges
with the following sizes:

1) An SU(3) magnetic charge identical to the one that if
alone would allow only the representations
of the {\underline group} $SU(3)/Z_{3}$, i.e. only to the representations
with triality t=0.

2) An SU(2) magnetic charge identical to the one that would allow
only representations of the {\underline{group}} $SU(2)/Z_{2}= SO(3)$,
i.e. only representations with integer weak isospin.

3) And finally, a U(1) weak hypercharge monopolic charge of a size
that if alone would allow only integer values of the weak
hypercharge half, i.e. of $y/2=$ integer.

These three magnetic charge contributions would, if alone,
not allow  the existence neither fermions, nor the Higgs bosons in the SM.
However, considering the phase of a quark or lepton field
along a little circle encircling the Dirac string for an
Olive's SM--monopole, one gets typically a phase rotation
from each of the three contributions to the magnetic charge.
The consistency condition to have the Dirac string
without visible effect is that these phase contributions
{\em together} make up a multiplum of $2\pi $.
It can be checked that the quark and lepton representations,
as well as the Weinberg--Salam Higgs boson representation,
lead to the full phase rotations which are indeed multipla
of $2\pi $.

Thus, as it was already mentioned above: if we imagine monopoles with each
of these contributions alone they would not allow the phenomenologically
observed quarks and leptons, nor the Higgs bosons.

In going to the FRGG--model we can, without problem,
postulate one Olive--monopole for each proto--family
since the proto--family representations are just
analogous to the ones in the SM.

Considering the Olive--monopoles condensation --- thereby causing a
confinement--deconfinement phase transition --- for the different families,
we assume that as long as we consider only, say,
the SU(3)--coupling to cause the phase transition, the
Olive--monopole functions as if it is the SU(3)--monopole
consistent only with the representations of $SU(3)/Z_3$.
But this is just what gives the phase transition couplings derived
with help of Eq.(\ref{5x}) together with Eq.(\ref{4x}).
Similarly, it is relatively easy to see that the use of the
Olive--monopole for all the (family) gauge groups
SU(3), SU(2), U(1) leads to the phase transition couplings
obtained by combining Eqs.(\ref{4x}) and (\ref{5x}).

\subsection{AntiGUT prediction of the fine structure constants near
the Planck scale}

According to AntiGUT \ct{2c}-\ct{2k}, the point where
FRGG undergoes spontaneous breakdown to its diagonal subgroup (isomorphic to
SMG) takes place at $\mu_G\sim 10^{18}$ GeV. This case is shown in Fig.2
demonstrating the evolution of the inverses $\alpha_{Y,2,3}^{-1}(\mu)$.  Here
$\alpha_Y\equiv \frac{3}{5} \alpha_1$ and:
\begin{equation}
   \alpha_Y^{-1}(\mu_{Pl})\approx 55.5.
                                                        \lb{83ya}
\end{equation}
A region of G--theory (where FRGGM works) also is shown in Fig.2.

If the point $\mu=\mu_G$ is very close to the Planck scale
$\mu=\mu_{Pl}$, then according to Eqs.(\ref{82y}) and (\ref{85y}), we have:
\begin{equation}
         \alpha_{1st\, fam.}^{-1}\approx
    \alpha_{2nd\, fam.}^{-1}\approx \alpha_{3rd\, fam.}^{-1}\approx
    \alpha_{Y,G}(\mu_G)\equiv \frac{\alpha_{Y}^{-1}(\mu_G)}{6}\approx 9.2,
                                                                  \lb{88y}
\end{equation}
which is almost equal to the value (\ref{50a}):
$\alpha_{crit,theor}^{-1}\approx 8$ obtained by the "Parisi improvement method"
and illustrated by Fig.4. This means that in the U(1)--sector of
AntiGUT we have $\alpha_{i,G}(\mu_G)$ near the (multiple)critical
point (MCP), and it is natural to assume that they obey the relations
(\ref{26z}) at least approximately:
\begin{equation}
   \alpha_{Y,G}^{-1}(\mu_G) : \alpha_{2,G}^{-1}(\mu_G) :
\alpha_{3,G}^{-1}(\mu_G) \approx 1 : \sqrt{3} : 3/\sqrt{2} = 1 : 1.73 : 2.12,
                                               \lb{26zA}
\end{equation}
For $\alpha_{Y,G}^{-1}(\mu_G)\approx 9.2$ (see Eq.(\ref{88y}))
the last equation gives the following result:
\begin{equation}
 \alpha_{Y,G}^{-1}(\mu_G) : \alpha_{2,G}^{-1}(\mu_G) :
\alpha_{3,G}^{-1}(\mu_G) = 9.2 : 15.9 : 19.5,             \lb{27z}
\end{equation}
which confirms the AntiGUT--MPM prediction \ct{17p}:
\begin{equation}
    \alpha_{Y,G}^{-1}(\mu_G)\approx 9.2\pm 1,
    \quad \alpha_{2,G}^{-1}(\mu_G)\approx 16.5\pm 1, \quad
    \alpha_{3,G}^{-1}(\mu_G)\approx 18.9\pm 1.
                                               \lb{28z}
\end{equation}
Using the relations (\ref{84y}), (\ref{85y}) and values (\ref{28z}),
we see that AntiGUT predicts:
\begin{equation}
    \alpha_Y^{-1}(\mu_G)\approx 55.5\pm 6,
    \quad \alpha_2^{-1}(\mu_G)\approx 49.5\pm 3, \quad
    \alpha_3^{-1}(\mu_G)\approx 57.0\pm 3,
                                               \lb{28zA}
\end{equation}
confirming the "experimental" evolutions of $\alpha_{Y,2,3}(\mu )$
shown in Fig.2.

If we believe in FRGGM existing near the Planck scale,
then the fall shown in Fig.5 also occurs near the Planck scale. Now the value
of $\alpha_{U(1)}^{-1}(\mu_{Pl}) = \alpha_{Y,G}^{-1}(\mu_{Pl})$ is 6 times
smaller than $\alpha_Y^{-1}(\mu_{Pl})$, and equal to the phase transition
(critical) point $\alpha_{U(1),crit}^{-1}$.

We can estimate the width of the effective potential jump $\Delta t$
at the Planck scale from Eq.(\ref{7x}) (see Section 7):
\begin{equation}
    \Delta t \approx - \frac{\Delta \alpha_Y^{-1}}{2\alpha_Y^{-1}}|_{t=0}.
                                              \lb{89y}
\end{equation}
Using the values (\ref{5x}) and (\ref{88y}), we obtain:
\begin{equation}
    \Delta t \approx \frac{\alpha_{Y,crit.Coul.} - \alpha_{Y,crit.conf.}}
      {2\alpha_{y,crot.conf}} \approx (3.7 - 9.2)/8 \approx - 0.7,
                                                    \lb{90y}
\end{equation}
Here we call $\alpha_{Y,crit.Coul.} \equiv \alpha_{Y,G}(\mu_G )
\approx 9.2$ --- the end point of the one--loop approximation straight line
in Fig.5b, whereas $\alpha_{Y,crit.conf.}\approx 3.7$ is the beginning
of the confinement phase.

Eq.(\ref{90y}) means that the one--loop approximation, shown in Fig.2 for
$\alpha_Y^{-1}(\mu)$, works up to the value

\begin{equation}
     \mu_G\approx \mu_{Pl}e^{- 0.35}\approx 0.7\mu_{Pl}\approx 8.5\cdot
                            10^{18}\,{\mbox{GeV}}.      \lb{91y}
\end{equation}
As a consequence, we see that in the AntiGUT approach
there is a sharp jump of $\alpha_Y^{-1}(\mu )$ from its one--loop value
at $\mu=\mu_G$ to its phase transition (critical) value at $\mu=\mu_{Pl}$.

\section{The possibility of the Grand Unification Near the Planck Scale}

In AntiGUT the FRGG breakdown was considered at $\mu_G\sim 10^{18}$ GeV.
It was a significant point for MPM.
In this case the evolution of the fine structure constants
$\alpha_i(\mu )$, also in the region $\mu > \mu_G$,
excludes the existence of the unification point (see Refs.
\ct{2a}-\ct{2k}, \ct{35}-\ct{38}, \ct{17p}-\ct{37}).

But the aim of this Section is to show that
we can see quite different consequences of the extension of SM to FRGGM
if $G$--group undergoes the breakdown to its diagonal subgroup (i.e. SM)
not at $\mu_G\sim 10^{18}$ {GeV}, but at $\mu_G\sim 10^{14}$ or $10^{15}$
{GeV}, i.e. before the intersection of $\alpha_{2}^{-1}(\mu)$ with
$\alpha_{3}^{-1}(\mu)$ at $\mu\approx 10^{16}$ GeV.

In fact, we here want to illustrate the idea that with monopoles we can
modify the running of the fine structure constants so much that unification
can be arranged without needing SUSY. To avoid confinement of the monopoles
at the cutoff scale we need FRGG or some replacement for it (see Section
4).

If we want to realize the behavior shown in Fig.5b for the function
$\alpha_{Y,FRGG}^{-1}(\mu )= \alpha_{Y,G}^{-1}$ near the Planck scale
we need the evolution of all $\alpha_i^{-1}(\mu)$ to turn away from
asymptotic freedom and that could be achieved by having more fermions,
i.e. generations, appearing above the $\mu_G$--scale.

Now we shall suggest that such appearance of e.g. more fermions above
$\mu_G$ is not at all so unlikely.

\subsection{Guessing more particles in FRGGM}

Once we have learned about the after all not so extremely simple SM, it is not
looking likely that the fundamental theory --- the true model of everything ---
should be so simple as to have only one single type of particle --- the
"urparticle" --- unless this "urparticle" should be a particle that can be
in many states internally such as say the superstring. We should therefore
not necessarily assume that the number of species of particles is minimal
anymore --- as could have been reasonable in a period of science where one
had only electron, proton and perhaps neutron, ignoring the photon exchanges
which bind the atom together, so that only three particles were really there.
Since now there are too many particles.

Looking at the problem of guessing the physical laws behind the Standard Model
in this light, we should rather attempt to guess a set of species of particles
to exist and the order of magnitudes of the numbers of such species which one
should find at the various scales of energy. Indeed, the historical learning
about the species of particles in Nature has rather been that physicists
have learned about many types of particles not much called for at first:
It is only rather few of particle types, which physicists know today that
have so great significance in the building of matter or other obviously
important applications that one could not almost equally well imagine a world
without these particles. They have just been found flavour after flavour
experimentally studies often as a surprise, and if for instance the charmed
quark was needed for making left handed doublets, that could be considered
as a very little detail in the weak interactions which maybe was not needed
itself.

These remarks are meant to suggest that if we should make our expectations
to be more "realistic" in the sense that we should get less surprised next
time when the Nature provides us with new and simingly not needed particles,
we should rather than guessing on the minimal system of particles seek to
make some more statistical considerations as to how many particle types
we should really expect to find in different ranges of energy or mass.

To even crudely attack this problem of guessing it is important to have
in mind the reasons for particles having the mass order of magnitudes.
In this connection, a quite crucial feature of the SM "zoo" is that except for
the Higgs particle itself, are mass protected particles, which have no mass
in the limit when the Higgs field has no vacuum expectation value (VEV).
In principle, you would therefore expect that SM particles have masses
of order of the Higgs VEV. Actually they are mostly lighter than that
by up to five orders of magnitude.

In the light of this mass protection phenomenon it would be really very
strange to assume that there should be no other particles than in SM
if one went up the energy scale and looked for heavier particles,
because then what one could ask: Why should there be only mass protected
particles? After all, there are lots of possibilities for making vector
coupled Dirac particles, say. It would be a strange accident if Nature
should only have mass protected particles and not a single true Dirac
particle being vector coupled to even the weak gauge particles. It is
much more natural to think that there are at higher masses lots of different
particles, mass protected as well as not mass protected. But because we
until now only could "see" the lightest ones among them, we only "saw"
the mass protected ones.

In this light the estimate of how many particles are to be found with
higher masses should now be a question of estimating how many particles
turn out mass protected and getting masses which we can afford to "see"
today.

We are already in the present article having the picture that as one goes
up in the energy scale $\mu $ there will be bigger and bigger gauge group,
which will thus be able to mass protect more and more particle types.
Each time one passes a breaking scale at which a part of the gauge group
breaks down --- or thought the way from infrared towards the ultraviolet:
each time we get into having a new set of gauge particles --- there will
be a bunch of fermions which are mass protected to get --- modulo small
Yukawa couplings --- masses just of the order of magnitude of the Higgs
scale corresponding to that scale of diminishing of the gauge group.

If we for example think of the scale of breaking of our FRGG down to its
diagonal subgroup, then we must expect that there should be some
fermions just mass protected to that scale. However, these particles
should be vector coupled w.r.t. the SM gauge fields, and only mass protected
by the gauge fields of FRGG, which are not diagonal. We would really like
to say that it would be rather strange if indeed there were no such
particles just mass protected to that scale.

\subsection{Quantitative estimate of number of particles in FRGGM}

We might even make an attempt to perform a quantitative estimate of how
many particle species we should expect to appear when we pass the scale
$\mu = \mu_G$ going above $\mu_G $. Of course, such an estimate can be
expected to be very crude and statistical, but we hope that anyway it would be
better than the unjustified guess that there should be nothing, although this
guess could be in some sense the simplest one.

Since it is going to be dependent on the detailed way of arguing, we should
like to make a couple of such estimates:

1) The first estimate is the guessing that, in analogy with the type of
particle combinations which we have had in FRGGM already, we find the
particles grouped into families which are just copies of the SM families.
If only one of the gauge group families is considered, then two others
are represented trivially on that family. We shall, however, allow that
these families can easily be mirror families, in the sense that they have
the weak doublets being the right handed particles and actually every
gauge quantum number parity are reflected. But, by some principle, only
small representations are realized in Nature and we could assume away
the higher representations. Now let us call the number of mirror plus
ordinary families which are present above the scale $\mu_G$ of the breakdown
of FRGG to the diagonal subgroup as $N_{fam,tot}$, and assume that we have
a statistical distribution as if these families or mirror families had
been made one by one independently of each other with a probability of 50\%
for it being a mirror family and 50\% for it being an ordinary left handed
one. The order of the number of families survived under the scale $\mu_G$
should then be equal to the order of the difference between two samples
of the Poisson distributed numbers with average $N_{fam,tot}/2$. In fact,
we might consider respectively the number of mirror families and the number
of genuine (left) families as such Poisson distributed numbers. The excess
of the one type over the other one is then the number of low energy scales
surviving families, and the physicists living today can afford to see them.
It is well known that crudely this difference is of the order of
$\sqrt{N_{fam,tot}}$.  But we know that this number of surviving families
has already been measured to be 3, and so we expect that
$\sqrt{N_{fam,tot}}=3$, what gives $N_{fam,tot}\approx 9$.
This would mean that there are 6 more
families to be found above the diagonal subgroup breaking scale $\mu_G$.

2) As an alternative way of estimating, we can say very crudely that the
fermions above the scale $\mu_G$ could be mass protected by 3 times as
many possible gauge quantum numbers, as far as there are three families
of gauge boson systems in our FRGGM. If we use the "small representations
assumption", then going from $\mu > \mu_G$ to $\mu < \mu_G$, two of three
fermions loose their mass protection, and these two fermions obtain masses
of order of the scale $\mu_G$. But this means that 1/3 of all fermions
survive to get masses below the scale $\mu_G$ and become "observable in
practice". Again we got that there should be two times as many particles
with masses at the diagonal subgroup breaking scale $\mu_G$ than at the
EW scale. We got the same result by two different ways. Notice though
that in both case we have used a phenomenologically supported assumption
that Nature prefers very small representations. In fact, it seems to be true
that the SM representations are typically the smallest ones allowed by the
charge quantization rule:
\be
     \frac{Y}{2} + \frac{d}{2} + \frac{t}{3} = 0 \quad ({\mbox{mod}}\, 1),
                                 \lb{G1}
\ee
where $d$ and $t$ are duality and triality, respectively.

\subsection{The FRGGM prediction of RGEs. The evolution of
fine structure constants near the Planck scale.}

Let us consider now, in contrast to AntiGUT
having the breakdown of G--group at $\mu_G\sim 10^{18}$ GeV,
a new possibility of the FRGG breakdown
at $\mu_G\sim 10^{14}$ or $10^{15}$ GeV (that is, before the
intersection of $\alpha_{2}^{-1}(\mu )\,$ with $\alpha_{3}^{-1}(\mu )$,
taking place at $\mu \sim 10^{16}$ GeV in SM). Then in the region
$\mu_G < \mu < \mu_{Pl}$ we have three $SMG\times U(1)_{(B-L)}$ groups
for three FRGG families, as in Refs.\ct{2a}-\ct{2k}, \ct{11}-\ct{38},
\ct{20p}-\ct{24p}, \ct{17p}-\ct{37}.
In this region we have, according to the statistical estimates in
Subsection 9.2, a lot of fermions (we designate
their total number $N_F$, maybe different with $N_{fam.tot}$), mass
protected or not mass protected, belonging to usual families or to mirror
ones.

Also monopoles can be important in the vicinity of the
Planck scale: they can give essential contributions to RGEs for
$\alpha_i(\mu )$
and change the previously considered evolution of the fine structure
constants.

Analogously to Eq.(\ref{16F}), obtained in Ref.\ct{47},
we can write the following RGEs for
$\alpha_i(\mu)$ containing beta--functions for monopoles:
\begin{equation}
\frac {d(\log \alpha_i(\mu))}{dt} =
   \beta(\alpha_i) - \beta^{(m)}(\tilde \alpha_i),\quad\quad i=1,2,3.
                                                           \lb{G2}
\end{equation}
We can use the one--loop approximation for $\beta(\alpha_i)\,$ because
$\alpha_i$ are small, and the two--loop approximation for dual beta--function
$\beta^{(m)}(\tilde \alpha_i)$ by reason that $\tilde \alpha_i$ are not very
small.  Finally, taking into account that in the non--Abelian sectors of FRGG
we have the Abelian artifact monopoles (see Section 8), we obtain the following
RGEs:
\begin{equation}
   \frac {d(\alpha_i^{-1}(\mu))}{dt} = \frac{b_i}{4\pi } +
   \frac{N_M^{(i)}}{\alpha_i}\beta^{(m)}(\tilde \alpha_{U(1)}),  \lb{G3}
\end{equation}
where $b_i$ are given by the following values:
$$
   b_i = (b_1, b_2, b_3) =
$$
\begin{equation}
( - \frac{4N_F}{3} -\frac{1}{10}N_S,\quad
      \frac{22}{3}N_V - \frac{4N_F}{3} -\frac{1}{6}N_S,\quad
      11 N_V - \frac{4N_F}{3} ).                   \lb{G4}
\end{equation}
The integers $N_F,\,N_S,\,N_V$ are respectively the total numbers
of fermions, Higgs bosons and vector gauge fields in FRGGM considered in
our theory, while the integers $N_M^{(i)}$
describe the amount of contributions of scalar monopoles.

The Abelian monopole beta--function in the two--loop approximation is:
\begin{equation}
      \beta^{(m)}(\tilde \alpha_{U(1)}) = \frac{\tilde \alpha_{U(1)}}{12\pi }
      (1 + 3 \frac{\tilde \alpha_{U(1)}}{4\pi }).      \lb{G5}
\end{equation}
Using the Dirac relation (\ref{7}) we have:
\be
      \beta^{(m)} =
      \frac{\alpha_{U(1)}^{-1}}{48\pi }
      (1 + 3 \frac{\alpha_{U(1)}^{-1}}{16\pi }),      \lb{G6}
\ee
and the group dependence relation (\ref{25z}) gives:
\be
      \beta^{(m)} =
      \frac{{C_i\alpha_i}^{-1}}{48\pi }
      (1 + 3 \frac{{C_i\alpha_i}^{-1}}{16\pi }),
                                                     \lb{G7}
\ee
where
\begin{equation}
   C_i = (C_1, C_2, C_3) =
        (\frac{5}{3}, \frac{1}{\sqrt{3}}, \frac{\sqrt{2}}{3}).
                                                          \lb{G8}
\end{equation}
Finally we have the following RGEs:
\begin{equation}
   \frac {d(\alpha_i^{-1}(\mu))}{dt} = \frac{b_i}{4\pi } +
   N_M^{(i)} \frac{{C_i\alpha_i}^{-2}}{48\pi }
      (1 + 3 \frac{{C_i\alpha_i}^{-1}}{16\pi }),         \lb{G9}
\end{equation}
where $b_i$ and $C_i$ are given by Eqs.(\ref{G4}) and (\ref{G8}), respectively.

In our FRGG model:
\begin{equation}
      N_V = 3,                       \lb{G10}
\end{equation}
because we have 3 times more gauge fields
in comparison with the SM ($N_{fam}=3$).

As an illustration of the {\em possibility} of unification in an
FRGG--scheme with monopoles we propose some --- strictly speaking
adjusted --- parameter choices within the likely ranges already
suggested.

In fact, we take the total number of fermions
$N_F=2N_{fam.tot}$ (usual and mirror
families), $N_{fam.tot}=N_{fam}N_{gen}=3\times 3=9$ (three SMG groups
with three generations in each group), we have obtained (see Fig.6)
the evolutions of $\alpha_i^{-1}(\mu )$ near the Planck scale
by numerical calculations for $N_F=18$, $N_S=6$, $N_M^{(1)}=6$,
$N_M^{(2,3)}=3$ and the following $\alpha_i^{-1}(\mu_{Pl})$:
\begin{equation}
   \alpha_1^{-1}(\mu_{Pl})\approx 13,\quad\quad
\alpha_2^{-1}(\mu_{Pl})\approx 19,\quad\quad
\alpha_3^{-1}(\mu_{Pl})\approx 24,                  \lb{G10a}
\end{equation}
which were considered instead of Eqs.(\ref{7z}).

We think that the values $N_M^{(i)}$,  which we have used here, are in
agreement with Eqs.(\ref{86yA})--(\ref{4da}), and $N_S=6$ shows
the existence of the six scalar Higgs bosons breaking FRGG to SMG
(compare with the similar descriptions in Refs.\ct{35}-\ct{38}).

Fig.6 shows the existence of the unification point.

We see that a lot of new fermions in the region
$\mu > \mu_G$ and monopoles near the Planck scale
change the one--loop approximation behaviour of
$\alpha_i^{-1}(\mu)$ in SM. In the vicinity of the Planck scale these
evolutions begin to decrease, approaching the phase
transition (multiple critical) point at $\mu = \mu_{Pl}$ what means
{\underline{the suppression of the asymptotic freedom in the non--Abelian
theories.}}

Here it is necessary to emphasize that these results do not depend
on the fact, whether we have in Nature lattice artifact monopoles or the
fundamental Higgs scalar particles having a magnetic charge (scalar
monopoles).

Fig.7 demonstrates the unification
of all gauge interactions, including gravity (the intersection of
$\alpha_g^{-1}$ with $\alpha_i^{-1}$), at
\begin{equation}
\alpha_{GUT}^{-1}\approx 27\quad {\mbox{and}} \quad
x_{GUT}\approx 18.4.                                     \lb{G11a}
\end{equation}

Here we can expect the existence of [SU(5)]$^3$
or [SO(10)]$^3$ unification (with/without SUSY).
Of course, the results obtained in this article are preliminary
and in future we are going to perform the spacious investigation of the
unification possibility.

But calculating the GUT--values for one family fine structure
constants in the case, considered in this paper, we have for i=5:
\begin{equation}
       \alpha_{GUT, one\,\, fam}^{-1} = \frac{\alpha_{GUT}}{3}\approx 9,
                                                             \lb{G11b}
\end{equation}
what corresponds to the Abelian monopole (for the SU(5)/$Z_5$ lattice
artifact) coupling with the total average monopolic fine structure constant
$\tilde \alpha_{eff}$ and the "genuine" monopole fine structure constant
$\tilde \alpha_{genuine}$ as defined in Ref.\ct{22p} (see Section 5,
Eq.(81) of this reference), determined from Eq.(79) and the Dirac relation
by
\begin{equation}
        \alpha_N^{-1}=\frac{N}{2}\sqrt{\frac{N+1}{N-1}}\cdot 4\tilde
\alpha_{eff} = \frac{2N}{N-1}\tilde \alpha_{genuine}
                                                           \lb{G11c}
\end{equation}
leading to
\begin{equation}
    \tilde \alpha_{eff}= 9\cdot
\frac{2}{5\cdot 4}\sqrt{\frac{4}{6}} = 0.7,
                                        \lb{G11d}
\end{equation}
and
\begin{equation}
   \tilde \alpha_{genuine}\approx \frac{2\cdot 4}{4\cdot 5}\cdot 9
\approx 3.6.               \lb{G11e}
\end{equation}
This value $\tilde{\alpha}_{eff}$ suggests that we may apply crude perturbation
theory both for monopoles and charges ($\alpha_{eff}\approx0.35$) if accepting that in the region
(\ref{22x}).

Critical coupling corresponds (see (\ref{51A})) to
$\tilde \alpha_{eff, crit} = 1.20$ giving for SU(5)/$Z_5$:
\begin{equation}
   \alpha_{5,crit}^{-1}\approx \frac{5}{2}\sqrt{\frac{6}{4}}\cdot 4\cdot
1.20 \approx 14.5,               \lb{G11f}
\end{equation}
meaning that the unified couplings for $[SU(5)]^3$ are already of
confinement strength.

Within the uncertainties we might, however, consider from
(\ref{G11b}) the coupling strength $\alpha_{GUT}\approx 1/9$ as being
equal to the critical value $\alpha_{5,crit}\approx 1/14$ from
Eq.(\ref{G11f}).

If indeed the $\alpha_{GUT}$ were so strong as to suggest confinement
at the unification point, it will cause the problem that the
fundamental fermions in SU(5) representations would be confined
and never show up at lower energy scales.

Assuming the appearance of SUSY, we can expect to have sparticles at the
GUT-scale with masses:
\begin{equation}
                 M\approx 10^{18.4}\quad{\mbox{GeV}}.         \lb{G11}
\end{equation}
Then the scale $\mu_{GUT}=M$, given by Eq.(\ref{G11}), can be considered
as a SUSY breaking scale.

The unification theory with $[SU(5)]^3$--symmetry was suggested first
by S.Rajpoot \ct{40} (see also \ct{40c}).

Note that since our possibility of unification has the very
high scale (\ref{G11}), it allows for much longer proton lifetime
than corresponding models with more usual unification scales,
arround $10^{16}$ GeV. This is true not only for proton decay
caused by the gauge boson exchange, but also by the triplet
Higgs exchange, since then the mass of the latter also may be put
up in scale.

Considering the predictions of such a theory for the low--energy
physics and cosmology, maybe in future we shall be able to answer the question:
"Does the unification of [SU(5)]$^3$ or [SO(10)]$^3$ type (SUSY or not
SUSY) really exist near the Planck scale?"

\section{Discussion of Some Various Scenarios of Working MPP}

In the present article we have --- mostly, i.e.except for Subsection 8.2
--- taken the picture that the inverse fine structure constants
$\alpha_i^{-1}$ run very fast being smaller as $\mu$ gets bigger in the
interval $[\mu_G,\mu_{Pl}]$.
The old literature \ct{17p} and Subsection 8.2,
however, rather take it that the running of the fine structure constants
between $\mu_G$ and $\mu_{Pl}$ is minute. It could be taken, for instance
by the philosophy of Ref.\ct{17p}, as lowest order perturbation theory
gives it because the scale ratio logarithm $\log(\mu_G/\mu_{Pl})$ is
supposed so small that the details of $\beta$--functions are hardly of
any importance. This point of view is in disagreement with the
expectation of quick strong jump put forward in Subsection 8.2.
The argument for there having to be such a jump in $\alpha_i^{-1}$
just before $t=0$ (when $\mu=\mu_{Pl}$) is based on {\em the assumption
that there is a phase transition at the Planck scale as a function of $t$}.
In principle, however, the occurrence of the jump depends on hard
computations {\em and on (precisely) how big the running
$\alpha_i^{-1}(t)$ are when approaching the Planck scale}. Also
that depends again on what (matter) particles exist and influence
the $\beta$--function (monopoles, extra fermions above $\mu_G$, etc.).

We would like here to list some options for obtaining
MCP--agreement in one version or the other one combined
with pictures of associated matter:

1) The option of Ref.\ct{17p} is to use

1a) the MCP-I definition;

1b) the approximation of very little running between
$\mu_G$ and $\mu_{Pl}$.

In this interpretation MCP-I, we do not really think of phases
as a function of the scale, but only as a function of the bare
parameters, e.g. the bare fine structure constants. Rather we have to
estimate what are the bare (or the Planck scale) couplings at
the phase transition point conceived of as a transition point
in the coupling constant space, but not as a function of scale.
Here we may simply claim that the Parisi improvement approximation
(\ref{49}) is close to the calculation of the "bare" coupling.
The Parisi improvement namely calculates an effective coupling
as it would be measured by making small tests of the effective
action of the theory on a very local basis just around one plaquette.
Indeed, we expect that if in the lattice model one seeks to measure
the running coupling at a scale only tinily under the lattice scale,
one should really get the result to be the Parisi improved value
corrected by a tiny running only.

Since even in this option 1), corresponding to Ref.\ct{17p}, we
want a phase transition which we here like to interpret as due to
monopole condensation --- in an other phase though --- we need
to have monopoles. Thus, we get the problem of avoiding these
monopoles in the $\beta$--function except for an extremely
small amount of scales. In the phase in which we live it is
suggested that the monopoles are made unimportant in the
$\beta$--function below $\mu_G$ by being {\em confined} by the
Higgs field VEVs which break the FRGG down to the SMG (possibly
extended with $U(1)_{(B-L)}$). Since the monopoles, we need, are
monopoles for the separate family-gauge-groups and since
most of the latter are Higgsed at the $\mu_G$--scale, we
expect the monopoles to be confined
into hadron--like combinations/bound states by the Higgs
fields at the scale $\mu_G$. Thus if $\mu_G$ is close to
$\mu_{Pl}$ --- in logarithm --- there will be very little
contribution of the $\beta$--function running due to monopoles.

In such a picture even extra fermions, as suggested in
Subsection 9.1, and the SM particles will not give much
running between $\mu_G$ and $\mu_{Pl}$.

Now there seems to be a discrepancy with calculations
in the MCP-II approach (Refs.\ct{21p}-\ct{24p} and \ct{10s})
which gave the phase transition point what in (\ref{50a}) is called
$\alpha_{crit.lat}^{-1}\approx 5$. But now we must remember that this value
was calculated as a long
distance value, i.e. not a "bare" value of the fine structure
constant in the lattice calculations of Jersak et al. \ct{10s}.  Also
our Coleman--Weinberg type calculation of it \ct{21p}-\ct{24p} was rather a
calculation of the renormalised (or dressed) coupling than of a bare
coupling.

The disagreement between $\alpha_{crit.lat}^{-1}\approx 5$
(= the critical coupling) and the bare coupling in the picture 1
sketched here is suggested to be due to renormgroup running in
other phases caused by monopoles there.

In fact, we have in this picture 1 other phases --- existing somewhere
else or at some other time --- in which there is no
breaking down to the diagonal subgroup, as in our phase.
In such phases the monopoles
for the family groups can be active in the renormalisation group over
a longer range of scales provided they are sufficiently light.
Assuming that the Schwinger's renormalisation scheme is wrong and that
the running due to monopoles make the coupling weaker at the higher scales
than at the lower energies, it could be {\it{in the other phase}} a
value corresponding to $\alpha_{crit.lat}^{-1}\approx 5$ in Eq.(\ref{50a})
for some lower $\mu$, say at the mass of monopoles, while it is still the
Parisi improved value at the bare or fundamental (Planck) scale.

Preliminary calculations indicate that requiring a large positive value of
the running $\lambda(\mu)$ at the cut off scale, i.e. a large positive
bare $\lambda_0$ in a coupling constant combination at the phase transition
makes the value $\alpha_{crit.lat}\approx 0.2$ (or $\alpha_{crit}\approx
0.208$ given by Refs.\ct{22p}-\ct{24p})
run to a bare $\alpha$ rather close to the Parisi improvement phase
transition coupling value (\ref{49}).

In order to explain this picture we need to talk about at least three
phases, namely, two phases without the Higgs fields performing the breaking
the group G to the diagonal subgroup and our own phase. In one of these
phases the monopoles condense and provide the "electric" confinement, while
the other one has essentially massless gauge particles in the family
gauge group discussed. In reality, we need a lot of phases in addition
to our own because we need the different combinations of the
monopoles being condensed or not for the different family groups.

2) The second picture uses MCP-II, which interprets the phase transition
required by any MCP--version as being a phase transition as a function
of the scale parameter, $\mu$ or $t$, and the requirement of MCP-II is
that it occurs just at the fundamental scale, identified with the Planck
scale.

Since the gauge couplings, if their running is provided (perturbatively)
only by the SM particles, would need $\alpha_{crit}^{-1}\approx 9$ rather
than $\alpha_{crit}^{-1}\approx 5$, further $\beta$--function effects are
needed.

Once the couplings get --- as function of $t$ --- sufficiently strong,
then, of course, perturbation theory gets unjustified, however, and higher orders or
nonperturbative effects are important. Indeed, higher order seems to help
strengthening of the electric couplings approaching the Planck scale.
But most crucially it is argued that the very fact of finding a phase
transition at $\mu_{Pl}$, as postulated by MCP-II in itself ---
see Sections 7 and 9 --- suggests that there must be
a rather quick running just below $\mu_{Pl}$. This picture 2
has such a property because of the extra fermions --- or whatever ---
which bring the strength of the fine structure constants up
for $\mu > \mu_G$, so that nonperturbative and higher order effects can
take over and manage to realize MCP-II.

\section{Conclusions}

In the present paper we have shown that
the existence of monopoles in Nature leads to the consideration of
the Family Replicated Gauge Groups (FRGG) of symmetry
as an extension of the Standard Model in the sense that the use of
monopoles corresponding to the family replicated gauge fields can bring the
monopole charge down from the unbelievably large charge which it gets in
the simple SM, according to the Dirac relation.

We have considered the case when our (3+1)--dimensional space--time
is discrete and has a lattice--like structure. As a consequence
of such an assumption, we have seen that the lattice artifact monopoles
play an essential role in FRGGM near the Planck scale: then
these FRGG--monopoles give perturbative contributions to
$\beta$--functions for the renormalization group equations
written for both, electric and magnetic fine structure constants,
and change the evolution of $\alpha_i(\mu )$ in the vicinity of
the Planck scale.

We have shown that in FRGG--model monopoles can produce the phase
transition (multiple critical) point at the Planck scale, indicating the
realization in Nature of the Multiple Point Principle (MPP) by
D.L.Bennett and H.B.Nielsen \cite{17p}.

We have calculated the critical couplings
$\alpha_{i,crit}=\alpha_i(\mu_{Pl})$ using the results obtained previously
in the Higgs Monopole Model \ct{22p}-\ct{24p} where the lattice
artifact monopoles are approximated by the Higgs scalar fields.

Here it is necessary to point out the coincidence of the results,
obtained in this paper, if we have in Nature lattice artifact monopoles,
or the Higgs scalar magnetically charged fields.

We discussed two scenarios implementing Multiple Critical Point
(MCP) existence.

It was also considered that the breakdown of FRGG at $\mu_G\sim 10^{18}\,$
GeV leads to the AntiGUT by C.D.Froggatt and H.B.Nielsen \ct{2k} with the
absence of any unification up to the scale $\mu \sim 10^{18}$ GeV.

Finally we have investigated the case when the breakdown of FRGG
undergoes at $\mu_G\sim 10^{14}$ GeV and is accompanied by
a lot of extra fermions in the region $\mu_G < \mu < \mu_{Pl}$.
These extra fermions suppressing the asymptotic freedom of non-Abelian
theories lead, together with monopoles, to the possible existence of
unification of all interactions including gravity at
$\mu_{GUT}=10^{18.4}$ GeV and $\alpha_{GUT}^{-1}=27$.

We discussed the possibility of the existence of the family replicated
unifications $[SU(5)]^3$ or $[SO(10)]^3$ (SUSY or not SUSY).

We have given a special case of new type of
unification. The realistic family replicated
unification needs a serious program of investigations showing
the consequences for the low--energy physics and cosmology
what we plan to do in future.

\newpage

{\bf ACKNOWLEDGMENTS:}\\

We would like to express special thanks to DESY Theory Group and
W.Buchm{\" u}ller for hospitality and financial support.

0ne of the authors (L.V.L.) thanks the Institute of Theoretical
Physics of Heidelberg University and B.Stech for hospitality and financial
support. It is a pleasure to thank all participants of the seminar of
this Institute for useful discussions.

L.V.L. is grateful to Axel Goertz from Frankfurt/M for his hospitality
and help.

We are deeply thankful to C.Froggatt, R.Nevzorov and Y.Takanishi for fruitful
discussions and comments.

This work was supported by the Russian Foundation for Basic Research
(RFBR), project \\
$N^o\,$ 02-0217379.

H.B.N. thanks the Humboldt Stiftung and DESY very much for the Humboldt
Preis.

\newpage
\clearpage
\begin{figure}

\noindent\includegraphics[width=159mm, keepaspectratio=true]{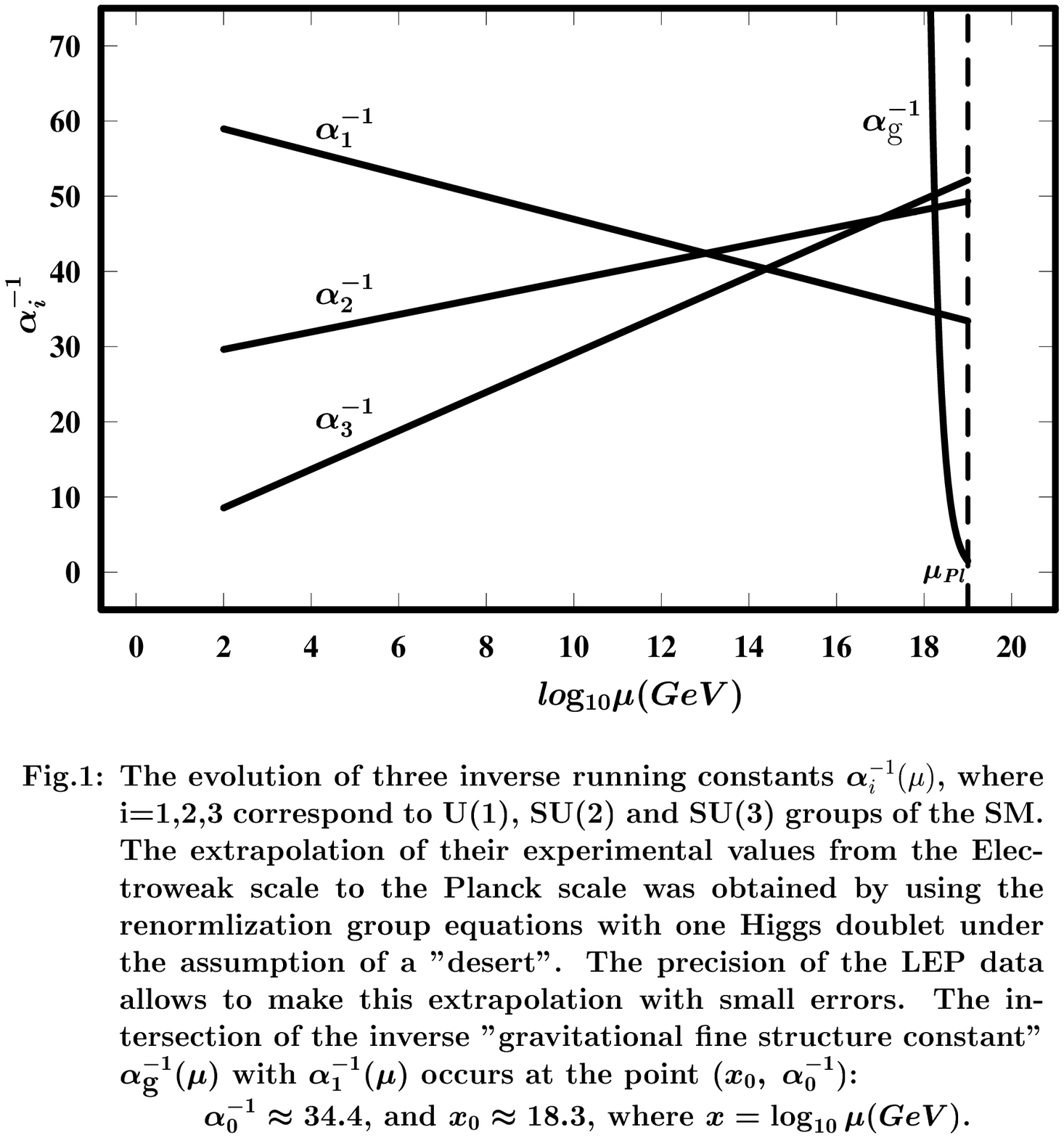}
\end{figure}

\newpage
\begin{figure}

\clearpage


\noindent\includegraphics[width=159mm, keepaspectratio=true]{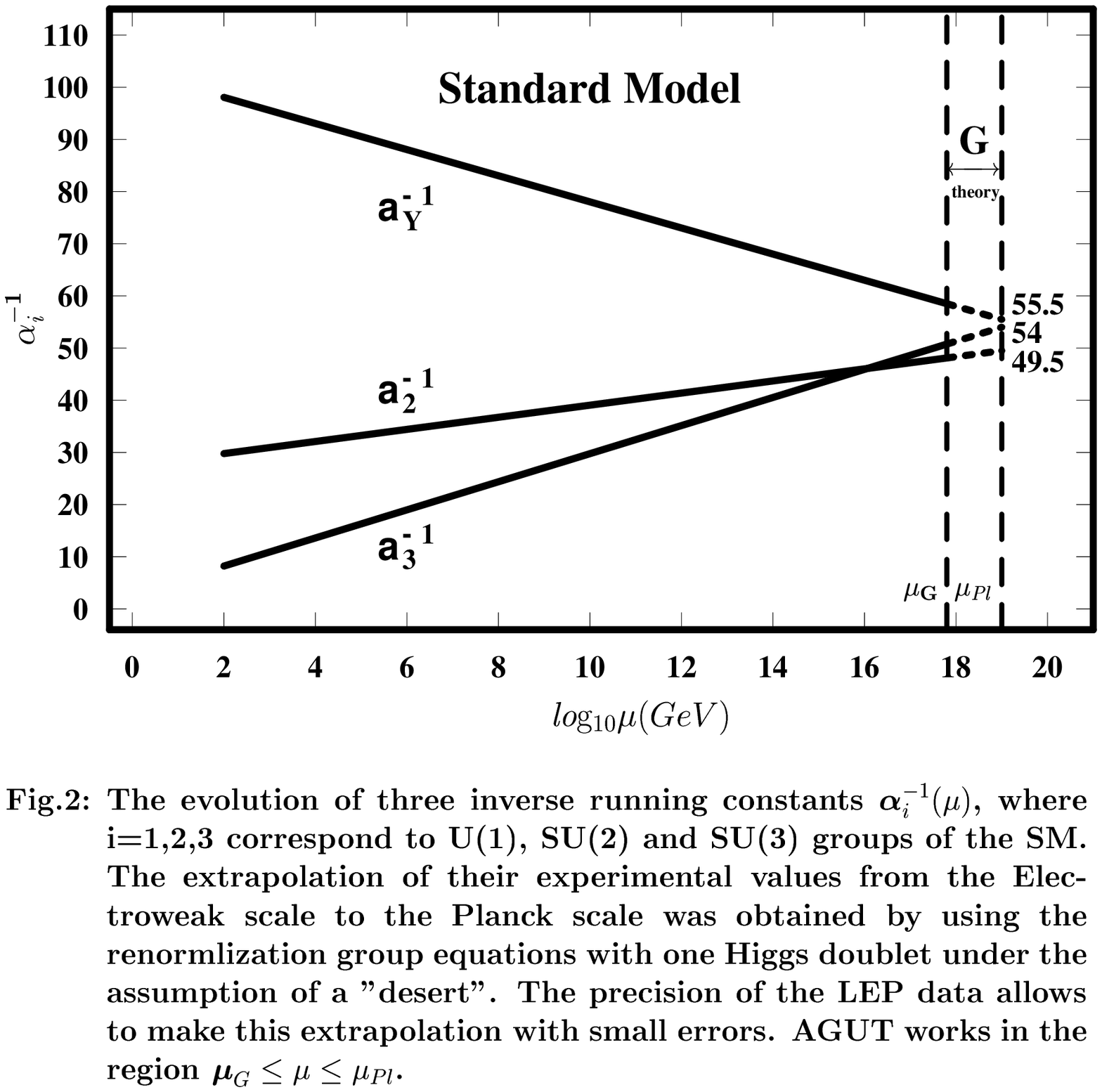}

\end{figure}

\newpage
\begin{figure}
\clearpage


\noindent\includegraphics[width=159mm, keepaspectratio=true]{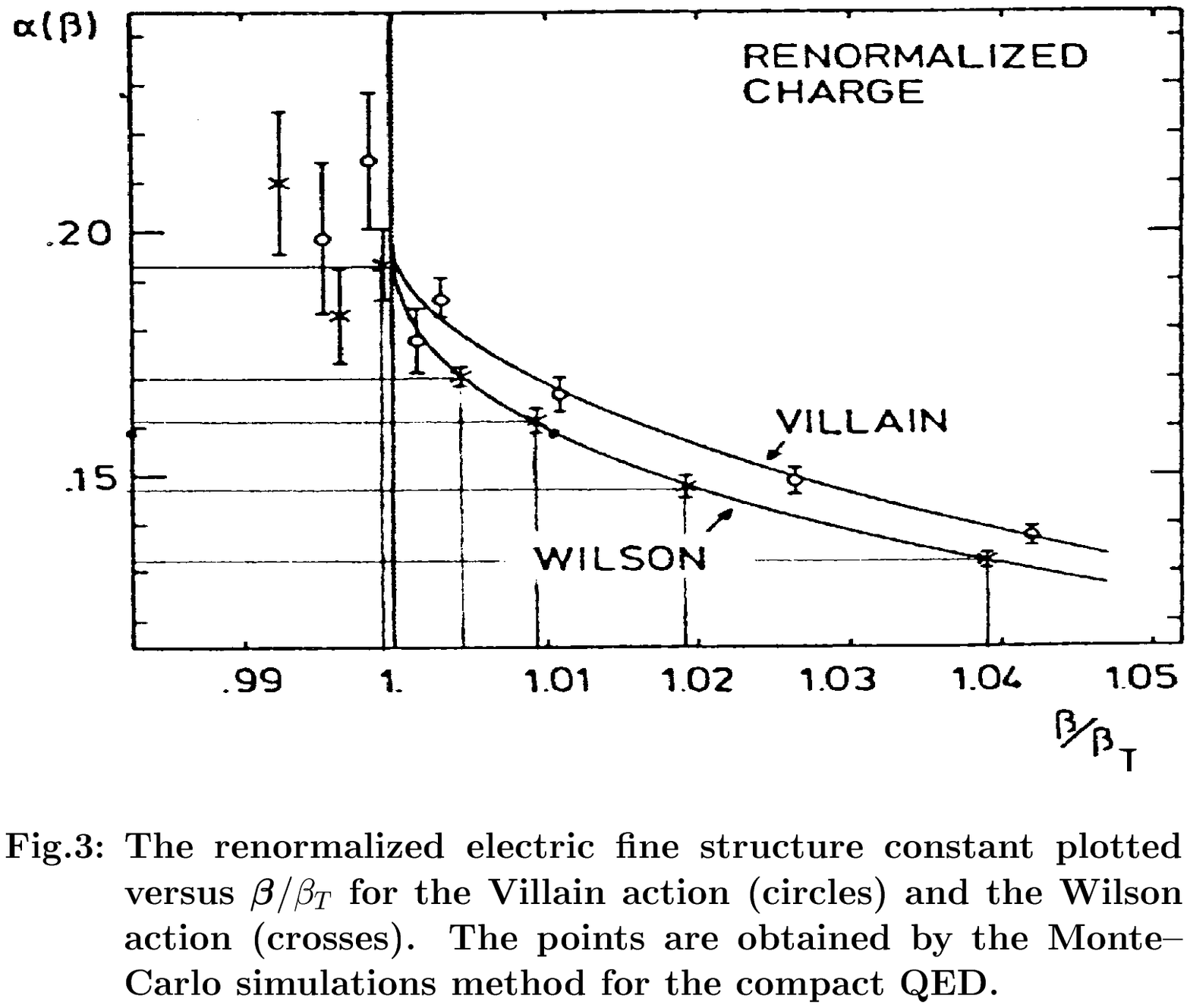}
\end{figure}

\newpage
\begin{figure}
\clearpage

\noindent\includegraphics[width=159mm, keepaspectratio=true]{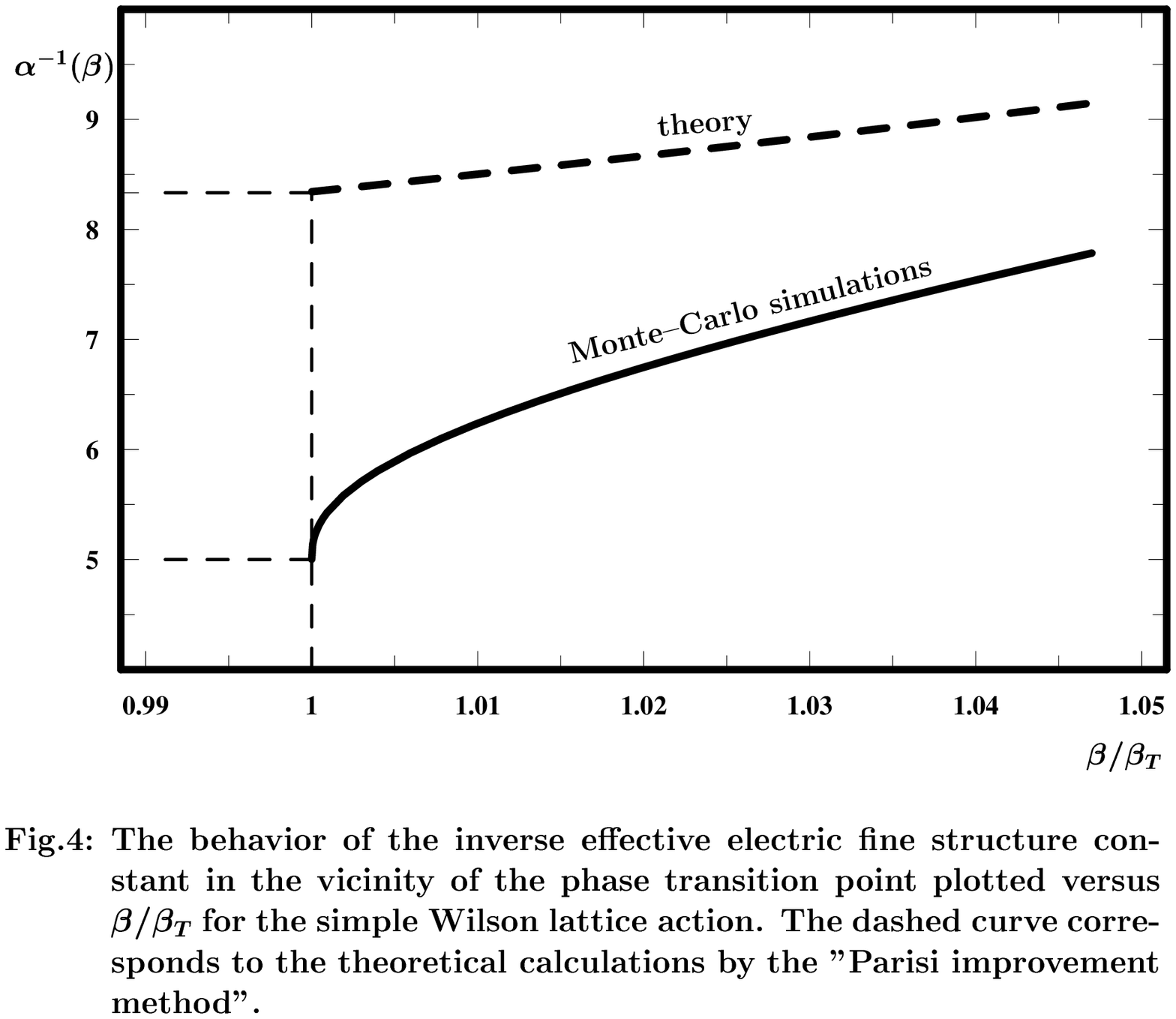}
\end{figure}

\newpage
\clearpage
\begin{figure}
\noindent\includegraphics[width=150mm, keepaspectratio=true]{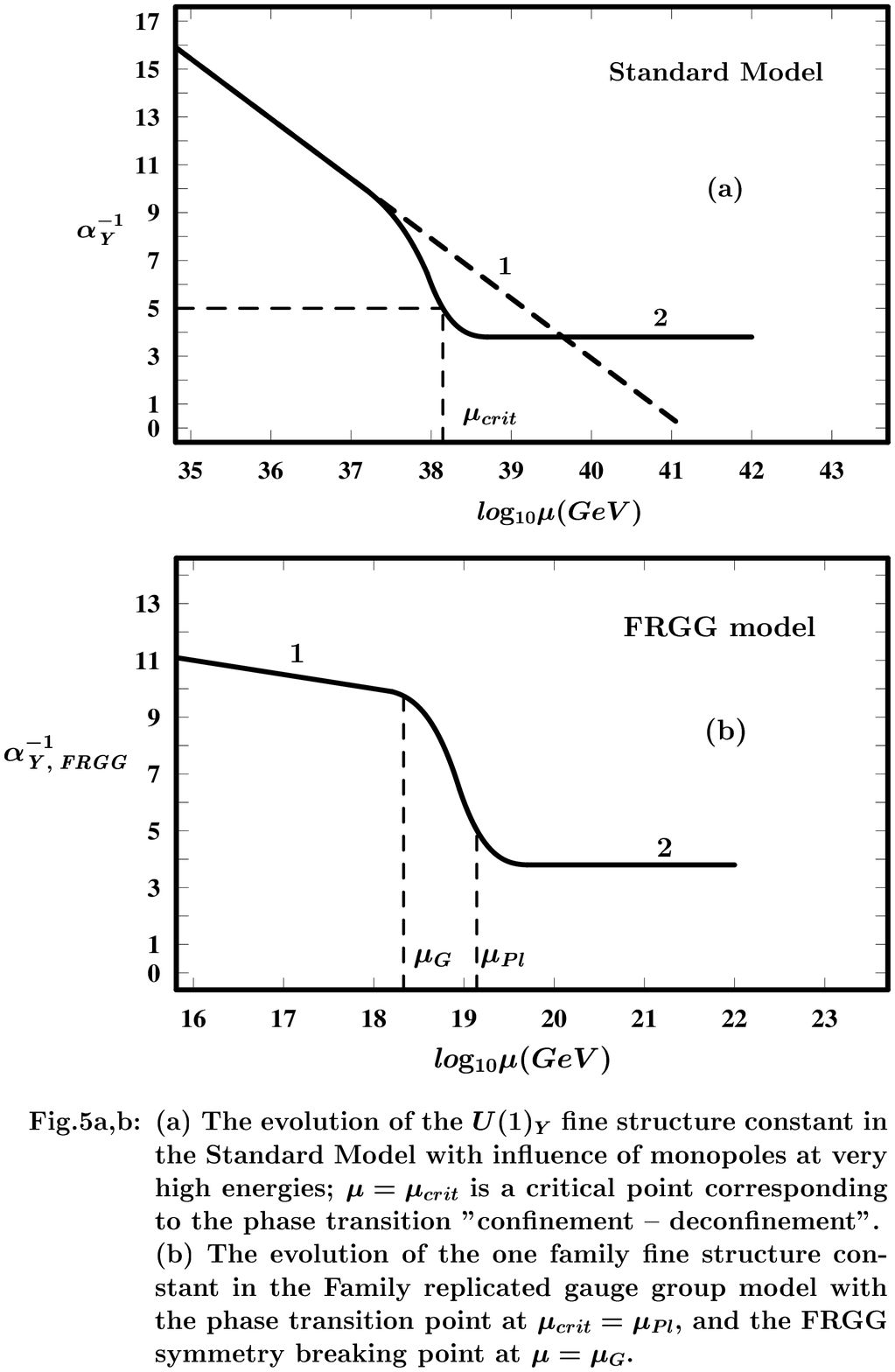}
\end{figure}

\newpage
\clearpage
\begin{figure}

\noindent\includegraphics[width=159mm, keepaspectratio=true]{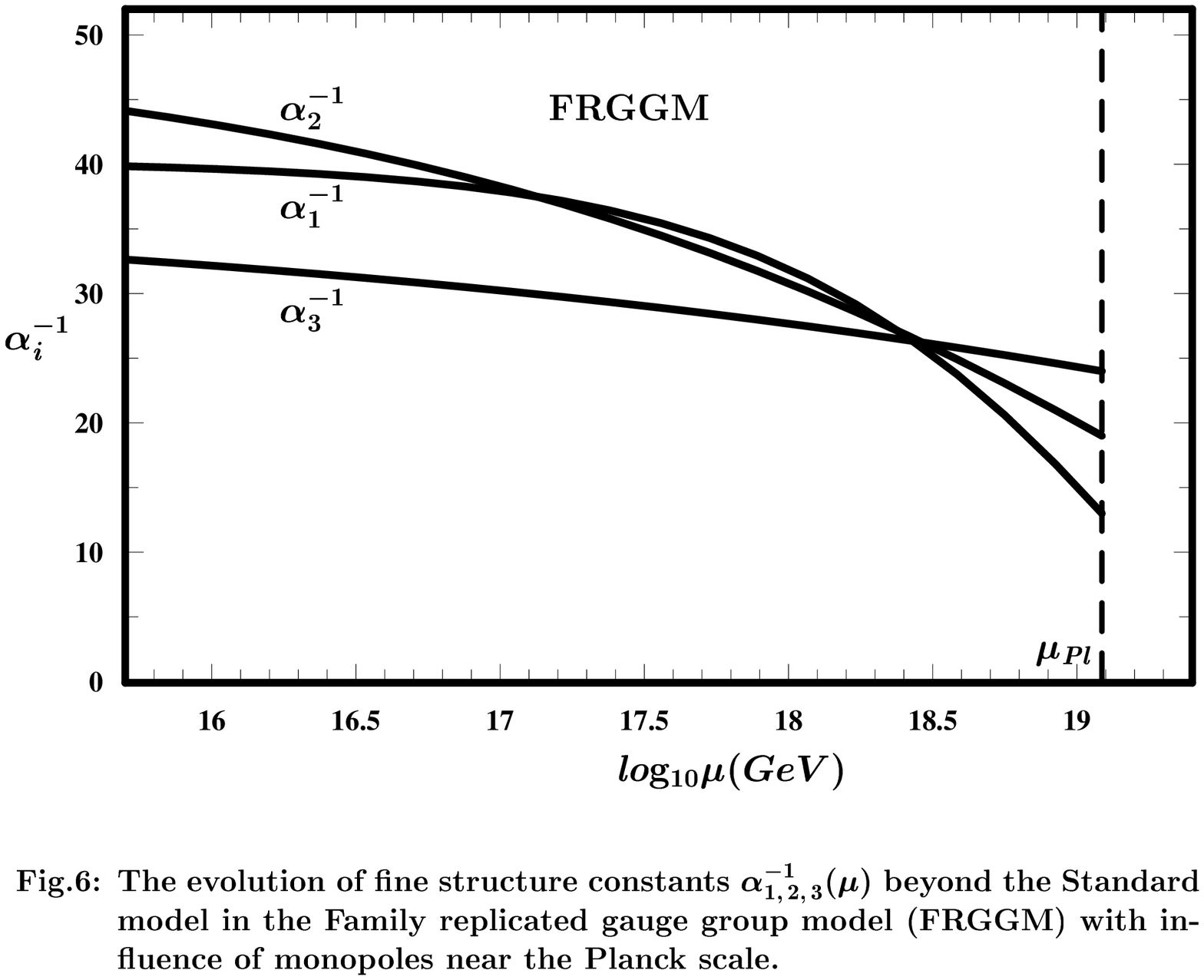}
\end{figure}

\newpage
\clearpage
\begin{figure}

\noindent\includegraphics[width=159mm, keepaspectratio=true]{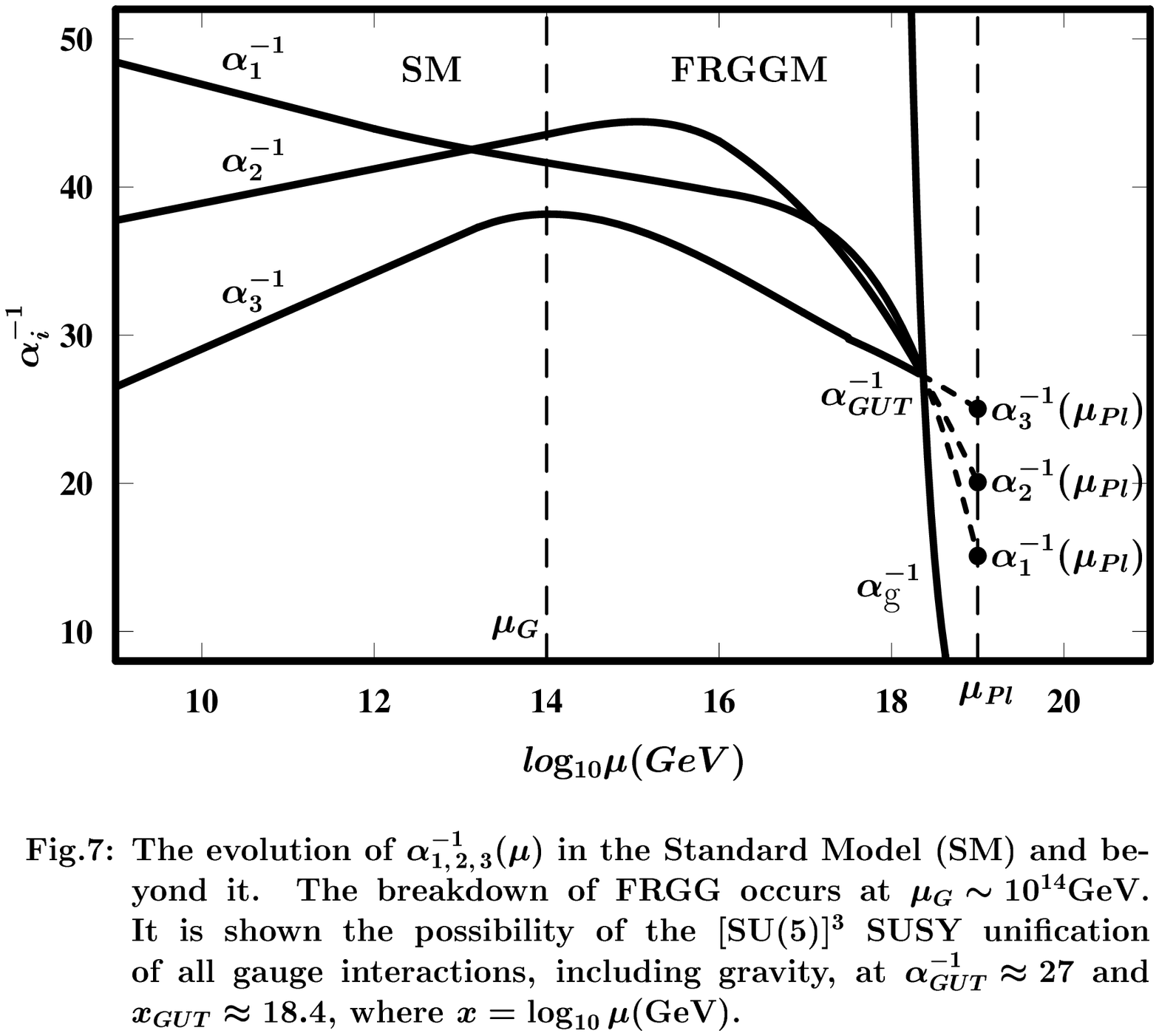}
\end{figure}

\end{document}